# Non-invasive imaging of Young's modulus and Poisson's ratio in cancers in vivo

Md Tauhidul Islam, Songyuan Tang, Chiara Liverani, Ennio Tasciotti, Raffaella Righetti*

*Abstract*—Objective: Alterations of Young's modulus (YM) and Poisson's ratio (PR) in biological tissues are often early indicators of the onset of pathological conditions. Knowledge of these parameters has been proven to be of great clinical significance for the diagnosis, prognosis and treatment of cancers. Currently, however, there are no non-invasive modalities that can be used to image and quantify these parameters in vivo without assuming incompressibility of the tissue, an assumption that is rarely justified in human tissues. Methods: In this paper, we develop a new method to simultaneously reconstruct YM and PR of a tumor and of its surrounding tissues, irrespective of the boundary conditions and the shape of the tumor based on ellipsoidal approximation. This new non-invasive method allows the generation of high spatial resolution YM and PR maps from axial and lateral strain data obtained via ultrasound elastography. The method was validated using finite element (FE) simulations and controlled experiments performed on phantoms with known mechanical properties. The clinical feasibility of the developed method was also demonstrated in an orthotopic mouse model of breast cancer. Results: Our results from simulations and controlled experiments demonstrate that the proposed reconstruction technique is accurate and robust. Conclusion: Availability of the proposed technique could address the clinical need of a non-invasive modality capable of imaging, quantifying and monitoring the mechanical properties of tumors with high spatial resolution and in real time. Significance: This technique can have a significant impact on the clinical translation of elasticity imaging methods.

## I. Introduction

Pathological changes typically alter the mechanical properties of tissues. In the case of many diseases such as cancer, atherosclerosis, fibrosis of the liver, etc., the tissues are reported to become harder as the disease develops [1], [2]. To describe the mechanical behavior of linear elastic tissues, at least two parameters are required - namely the Young's modulus (YM) and the Poisson's ratio (PR). YM is a mechanical parameter than can be used as a measure of the stiffness of a tissue. In homogeneous linear elastic solids, the YM is defined as the ratio of the applied axial stress and the resulting axial strain. In inhomogeneous solids, the YM can be obtained by dividing the local axial stress developed inside the solid due to the applied stress by the resulting local axial strain [3]. PR, on the other hand, provides a measure of the compressibility of the solid. In homogeneous linear elastic solids, the PR is defined as the negative ratio between the lateral strain and the axial strain. In inhomogeneous solids, the PR is related to the lateral to axial strain ratio in a more complex manner [3]. While many studies reported in the literature assume that biological tissues behave as linear elastic solids, there is a growing body of evidence indicating that poroelastic models may provide a more realistic description of the mechanical behavior of complex tissues than linear elastic models [4]–[6]. A poroelastic material is, by definition, compressible. Immediately after the application of the axial stress, a poroelastic tissue behaves as an incompressible solid with PR of $0.5$. Then, relaxation takes place inside the tissue during which dynamic processes occur, and the strain distributions inside the material undergo spatial and temporal changes. At steady state (also referred to as drained condition), the tissue behaves as a linear elastic solid. Therefore, YM and PR can still be used to quantify the stiffness and compressibility of a poroelastic tissue, as long as their measurements are computed in steady state conditions.

There are only a few non-invasive imaging modalities that are capable of generating YM maps of tissues in vivo and no available methods to non-invasively image the actual PR in complex biological tissues. Ultrasound elastography (USE) [7], ultrasound shear wave elastography (SWE) [8] and magnetic resonance elastography (MRE) [9] techniques have shown to be able to provide YM images, under the assumption that the tissue behaves as a linearly elastic incompressible solid (i.e., with a PR of $0.5$ or close to $0.5$) [10]–[15]. Recent studies have demonstrated the feasibility of imaging the lateral-to-axial strain ratio, also referred to as effective PR (EPR), in tissues using elastography [16]–[18], but not the actual, underlying PR of the tissue.

Poroelastography is a new elastographic technique that aims at assessing the poroelastic behavior of tissues by analyzing the temporal and spatial distributions of the local axial and lateral strains (and related parameters) in the tissue while it is under compression [17], [19]. In an ultrasound poroelastography experiment, the tissue is compressed for a certain time interval while a series of radio frequency (RF) data is acquired, from which time-dependent axial and lateral elastograms are computed. Since at steady state the tissue behaves as a linear elastic material, from knowledge of the steady state axial and lateral strain distributions, it is possible to determine the local YM and PR of the poroelastic tissue using the formulations

Md Tauhidul Islam, Songyuan Tang and *R. Righetti are with the Department of Electrical and Computer Engineering, Texas A&M University, College Station, TX 77843 USA (e-mail: righetti@ece.tamu.edu), Chiara Liverani is with Osteoncology and Rare Tumors Center, Istituto Scientifico Romagnolo per lo Studio e la Cura dei Tumori (IRST) IRCCS, Meldola, Italy, Ennio Tasciotti is with Center of Biomimetic Medicine, Houston Methodist Research Institute, 6670 Bertner Avenue, Houston, TX 77030, USA.

for linear elasticity theory.

There are two main approaches that can be used for reconstructing the YM distribution in tissues - a direct approach and an iterative approach. In the direct approach, a partial differential equation developed using the equations of equilibrium for linear elastic solids is used for estimating the YM [20]–[22]. The limitation of the direct approach is that it implicitly assumes continuity of the stress and strain in the tissue. Therefore, it is not directly applicable to cases where the distribution of the YM can vary sharply such as at the interface of a tumor and surrounding tissue.

The iterative methods utilize forward and backward solutions of the differential equations of equilibrium for linear elastic solids and attempt to minimize their differences [10], [11]. Generally, the iterative methods are more robust than the direct approaches but have other limitations. Firstly, these methods are computationally intensive [23], [24]. Secondly, they require a regularization term, which is often difficult to choose. The regularization parameter is used to reduce the noise and preserve the contrast of the reconstructed YM image. Inability to select a proper value of the regularization parameter can result in incorrect and noisy estimates of YM.

In most of the works pertinent to medical elasticity imaging retrievable in the literature, the YM of the tissue is reconstructed with two fundamental assumptions: 1) that the tissue (tumor and surrounding tissue) behaves as a perfectly linearly elastic solid, and 2) that the tissue is incompressible or nearly incompressible [10]–[15], [25]. The first assumption allows these methods to estimate the YM of the tissue from knowledge of the instantaneous strain in response to the applied compression. Based on the second assumption, the PR of the tissue, which is also needed to correctly estimate the YM of the tissue, is not estimated. Rather, it is assumed to be a given value, typically $0.495/0.499995/0.45$. [10]–[15], [25]. In regard to the first assumption, it is now widely believed that tissues can be more realistically represented using poroelastic models instead of linearly elastic models [5], [26], [27]. Thus, their strain response under loading varies with time. In that case, the YM and PR should be determined by the strain response at steady state, when the material is fully relaxed [4], and not by the instantaneous response. In fact, the YM estimated from the instantaneous strain in soft tissues can be significantly higher (2-4 times) than the true YM value as shown in Bayat et al. [28]. In regard to the second assumption, it has been demonstrated by a number of prominent studies that the PR of tissues (including tumors) may be significantly lower than $0.495$. In the works of Stylianopoulos et al. [26], Mpekris et al. [29] and Fung [30], the PR of normal tissue was assumed $0.2$ and that of cancer was assumed $0.2$ (compressible)/$0.45$ (incompressible) in works of Stylianopoulos et al. [26], Roose et al. [31] and Netti et al. [6]. Recently, Nia et al. [4] assumed a PR value for the soft tissue and tumor of $0.1$ to compute the residual stress inside the tumor. In some other works [32]–[35], the authors reported or used values of PR for the soft tissue ranging between $0.3$ and $0.45$. Given the broad range of PR values for soft tissue and tumors that has been reported in the literature, the assumption that the PR is constant and equal to $0.5$ or a value to close to $0.5$ is not only unrealistic but also can lead to incorrect reconstructed mechanical parameters values. Accurate determination of the PR is crucial to obtain accurate estimates of YM. In addition, a correct knowledge of the YM and PR is essential for the quantification of other poroelastic parameters such as vascular permeability and interstitial permeability, which are known to be of great clinical value [27]. Finally, it may be reasonable to expect that the PR itself may change with the onset of many diseases [17] as it is directly related to the compressibility of the tissue, and this information could prove useful clinically.

In addition to the tissue incompressibility assumption, most YM reconstruction methods retrievable in the literature present a series of limitations. The estimation of the mechanical properties of tumors is inherently a three-dimensional problem. While a few three dimensional YM reconstruction methods have been reported in the literature [15], [36], in most of the prior YM reconstruction studies, the models are two-dimensional and based on the common assumption of plane strain/plane stress [10], [11], [37]–[40]. Most of retrievable methods assume specific boundary conditions such as total uniformity of the background, stress-free lateral boundaries etc., which are rarely true in complex tissue environments. Most of these methods perform well for tumors of specific shapes such as disk (2D)/sphere (3D) [10], [36], [39], [40] but have poorer performance in tumors of other shapes such as ellipse. In case of soft tumor or tumor of high YM contrast, the strain inside the tumor changes small for a large change in the tumor YM. Therefore, estimation of YM in these cases requires higher sensitivity of the method to strains inside the tumor. As the available methods rely only on the change of axial strain neglecting the change in the lateral strain, they cannot perform efficiently when the YM contrast between the tumor and normal tissue is larger than $10$ or when the tumor is softer than the background [15], [36], [40]. In many cases, the tumor is assumed to be very small so that certain ratios such as ratios of sample-radius-to-tumor-radius, compressor-radius-to-tumor-radius and distance between applied force and tumor-to-tumor-radius are greater than a predefined value [36], [39]. Determination of heterogeneous distribution of YM inside the tumor and normal tissue is another challenge [41]. Most of these methods fail to reconstruct the YM accurately in case of non uniform axial compression, which occurs frequently in elastography experiments.

In this paper, we present a three-dimensional method that allows reconstruction of both YM and PR based on Eshelby's inclusion formulation [3], [42]. Our proposed method overcomes the aforementioned limitations of current YM reconstruction methods. It allows simultaneous quantification and imaging of the YM and PR in both a tumor and surrounding tissue irrespective of the complex boundary conditions and/or the shape of the tumor and for a wide range of tumor/background YM contrasts $(0.1 - 50)$. In our approach, the tumor and normal tissues are assumed to behave as poroelastic materials, and the YM and PR are reconstructed from knowledge of the strain responses at steady state. The proposed method is based on a cost function minimization technique, and the cost function is developed utilizing the formulations of eigen strain described in the works of Eshelby [3] and Mura [42].

In the proposed technique, no assumption on the boundary conditions is imposed, and the reconstructed YM and PR of the tumor depend only on the known value of applied stress, the strains inside the tumor and in the background and the geometry of the tumor. Thus the estimated parameters are less influenced by inhomogeneities present in the normal tissue or boundary conditions created by the experimental protocol and cancer environment. For these reasons, our method is less prone to error when the applied compression is not uniaxial or the tumor is not small compared to the background. Our method is sensitive to changes in both axial and lateral strains inside the tumor, which enables us to estimate Young's modulus of both soft and hard tumors accurately. In this study, we demonstrate that when the complex shapes of the tumor are approximated with an ellipse, such approximation introduces small errors in the reconstructed YM and PR of the tumor. As our technique reconstructs the YM and PR at each pixel inside the tumor simultaneously and independently, it can accurately reconstruct the heterogeneous distribution of the YM. The proposed method is validated both with simulations and controlled experiments for a large number of samples and tested in animal experiments in vivo. Our results are compared with those obtained using two other YM reconstruction methods available in the literature, where incompressibility of the tissues is assumed. The results show the presence of significantly smaller errors in the estimated YM when the proposed method is used in comparison to the previously proposed methods. These results not only prove the superiority of the proposed technique in terms of accuracy, resolution and robustness with respect to the existing methods but also demonstrates that the assumption of tissue incompressibility typically made in elastography applications can lead to significant errors in the reconstructed YM of the tissue.

## II. Proposed Method

The local stress and strain inside and outside an inclusion due to the remote stress have been determined by Eshelby [3] using the superposition principle and Green's function. The remote stress is the applied stress that creates a uniform stress over the entire background. This was done using a virtual experiment, which is summarized in Fig. S17. Based on the virtual experiment, the strain and stress inside the inclusion can be written as [3], [15]

$$\epsilon = \epsilon^0 + S : \epsilon^*, \quad (1)$$

$$\sigma = \sigma^0 + C^0 \cdot [S - I] : \epsilon^*, \quad (2)$$

where $\epsilon^0$ is the remote strain, $\epsilon^*$ is the eigenstrain, $\sigma^0$ is the remote stress, $C^0$ is the stiffness tensor of the background, $I$ is the identity tensor and $S$ is the Eshelby's tensor. $\epsilon^0$, $\epsilon^*$ and $\sigma^0$ are vectors of three components (axial, lateral and elevational). The relationship between the remote stress $\sigma^0$ and $\epsilon^0$ can be expressed as

$$\sigma^0 = C^0 : \epsilon^0. \quad (3)$$

The Eshelby's tensor $S$ is a function of the geometry of the inclusion and the PR of the background.

In eqs. (1) and (2), the eigenstrain can be written as ([42] eq. 22.13)

$$\epsilon^* = (S + A)^{-1} : (B : \epsilon^t - \epsilon^0), \quad (4)$$

where $\epsilon^t$ is a prescribed eigenstrain (zero for our current problem), and the fourth-order mismatch stiffness tensors A and B can be defined as

$$A = [C - C^0]^{-1} \cdot C^0, \; B = [C - C^0]^{-1} \cdot C, \quad (5)$$

where $C$ is the stiffness tensor in the inclusion. The expression of $A$ is relevant to our problem. $A$ is determined in eq. 10 in section 11 of supplementary information.

Let us indicate $\epsilon^*$ in eq. (1) as $\epsilon_1^*$ and $\epsilon^*$ in eq. (4) as $\epsilon_2^*$. In the expression of $\epsilon_1^*$, only the Eshelby's tensor $S$ is involved. This requires knowledge of the tumor (inclusion) geometry and the PR of the normal tissue (background). In the expression of $\epsilon_2^*$, the YM and PR of the tumor and normal tissues are involved.

A cost function can be defined as

$$J(E_i, \nu_i) = (J_1(E_i, \nu_i))^2 + (J_2(E_i, \nu_i))^2 + (J_3(E_i, \nu_i))^2, \quad (6)$$

where

$$J_1(E_i, \nu_i) = \epsilon_1^*(1) - \epsilon_2^*(1), \quad J_2(E_i, \nu_i) = \epsilon_1^*(2) - \epsilon_2^*(2),$$
$$J_3(E_i, \nu_i) = \epsilon_1^*(3) - \epsilon_2^*(3) \quad (7)$$

and by minimizing this cost function $J$, we can obtain the YM ($E_i$) and PR ($\nu_i$) of the tumor. Let us assume that the geometry of the tumor is axisymmetric, i.e., the dimension of the tumor is same along lateral and elevational direction. In this case, the lateral and elevational components of strains are equal and we can write the cost function as

$$J(E_i, \nu_i) = (J_1(E_i, \nu_i))^2 + (J_2(E_i, \nu_i))^2. \quad (8)$$

The YM and PR of the normal tissue can be determined by using eq. (3). The expressions of $\epsilon_1^*$ and $\epsilon_2^*$ for elliptic (prolate, oblate) and spherical tumor (inclusion) are shown in supplementary information (sections 12 and 16). The expressions of the Eshelby's tensor $S$ for cylindrical, flat elliptic, penny-shaped tumors are given in supplementary information (sections 13, 14 and 15). Using these $S$ in equations of $\epsilon_1^*$ and $\epsilon_2^*$ for elliptic tumor, $\epsilon_1^*$ and $\epsilon_2^*$ for these shapes can be determined (and therefore YM and PR).

## III. Simulations

The methods of FE and ultrasound simulations are discussed in supplementary information (sections 1 and 6). Eight samples of different shapes (Z1-Z8), nine samples of different inclusion/background YM contrasts (fixed inclusion/background PR contrast) (X1-X9), three samples with different boundary conditions (B1-B3), three samples with different YM heterogeneity percentages (H1-H3), four samples with different non-uniform loadings (R1-R4) and thirteen samples of different inclusion/background YM and PR contrasts (A-M) were simulated and analyzed. Details on the method to simulate different boundary conditions are given in supplementary information (section 2). The simulation methods used to create the heterogeneity of the YM distribution inside the tumor are discussed in supplementary information (section 3). The procedure to simulate non-uniform loading conditions is discussed in supplementary information (Section 4). Detail specifications of the samples used in simulations are added in supplementary information (section 1).

*A. Calculation of RMSE for the estimated YM and PR*

Calculation of percent root mean squared error (RMSE) for the estimated YM and PR of the inclusion from FE and ultrasound simulations was performed using the following formula [16].

$$\text{RMSE} = \sqrt{\frac{\sum_n^N (\Lambda_e(n) - \Lambda_t(n))^2}{N}} \times \frac{100 \times N}{\sum_n^N \Lambda_t(n)}, \quad (9)$$

where $\Lambda_e$ is the vectorized (reshaped from 2D to 1D) YM or PR of the inclusion from YM and PR images estimated by different methods and $\Lambda_t$ is the vectorized true YM or PR of the inclusion. $N$ is the total number of points inside the inclusion of the estimated YM or PR image.

## IV. EXPERIMENTS

*A. Controlled experiments*

For the controlled experiments, we used the breast phantom model 059 from Computerized Imaging Reference Systems (CIRS), Inc., Norfolk, VA, USA. As provided by the manufacturer, in this phantom, the YM of each inclusion mass is around 50 kPa, while the background has a YM of $20 \pm 5$ kPa [43]–[45]. The PR of both inclusions and background of this phantom is 0.5 [46]. The applied compression was monitored using a force sensor with a graphical user interface. A schematic of the setup for the controlled experiments is shown in Fig. S28. The axial and lateral strains were estimated using the pre- and post-compressed ultrasound radio frequency data acquired during the elastography experiment. A $5 \times 5$ pixels median filter was used on the axial and lateral strain elastograms obtained from the controlled experiments before reconstructing the YM and PR.

*B. In vivo experiments*

Experiments on nineteen mice with triple negative breast cancer cells injected in the mammary fat pad were carried out on a weekly basis for three consecutive weeks. The cancers were created at the Houston Methodist Research Institute by injection of the cancerous cells beneath the mouse's mammary fat pad [47]. In vivo data acquisition was approved by the Houston Methodist Research Institute, Institutional Animal Care and Use Committee (ACUC-approved protocol # AUP-0614-0033). Seven mice were kept untreated and twelve mice were treated by injecting them intravenously with one of the following drugs: 1. Epirubicin alone, 2. Liposomes loaded with Epirubicin and 3. Liposomes loaded with Epirubicin and conjugated with Lox antibody on the particle surface. The dose of each drug was 3 mg/kg body weight once a week. Prior to ultrasound data acquisition, each mouse was anesthetized with isoflurane. Each data acquisition session was 5 minutes long, and several RF data acquisitions could be performed during this period (for reliability purposes).

Elastography was carried out using a 38-mm linear array transducer (Sonix RP, Ultrasonix, Richmond, BC, Canada) with a center frequency of 6.6 MHz and $5 - 14$ MHz bandwidth. To compensate for the surface geometry as well as facilitate positioning the focus inside the superficial tumors, an aqueous ultrasound gel pad (Aquaflex, Parker Laboratories, NJ, USA) was placed between the compressor plate and the tumor. It should be noted that such use of gel pad does not change the stress distribution inside the sample significantly and thus does not change the estimated parameters. This has been proved in supplementary information (section 9). A force sensor (Tekscan FlexiForce) was inserted between the gel pads top surface and the compressor plate to record the applied force during the compression. Creep compression was performed in free-hand mode on the animals and monitored using the force sensor, with the duration of each compression being one minute. Duration of the experiment was selected based on the temporal behavior of the soft tissue and tumor reported in the literature [48] and to ensure that both the tumor and surrounding tissues reached steady state conditions. Ultrasound radio-frequency (RF) data acquisition was synchronized to the application of the compression. The sampling period of the data was set at 0.1 s. The axial and lateral strain data were calculated at steady state, when both the tumor and normal tissues behave as elastic materials [49]. The temporal curves of axial and lateral strains in a creep experiment are shown in Fig. S27, and the procedure to determine their values at steady state is discussed in supplementary information (section 23). An expert radiologist is employed to segment the in vivo axial strain elastograms in Matlab for determining the tumor areas.

*C. Calculation of applied stress*

FlexiForce OEM Development Kit manufactured by Tekscan, Inc., South Boston, MA, USA-02127 was employed to inspect and adjust the applied compression in both the controlled and in vivo experiments. A Microsoft Windows based interface software is provided with the sensor and can be used to observe and record the applied force. A temporal curve showing the applied compression in one of the in vivo experiments is reported in Fig. S18. The sensor used in the kit is model #A201, which senses a force range $0 - 4.4$ N in a scale of $0 - 255$. The diameter of the sensing area of the sensor is 9.53 mm. The sensing area is calculated as $7.1331 \times 10^{-5}$ m$^2$ ($A_r = \pi r^2$). The applied pressure in Pa is calculated using

$$\sigma_0 = \frac{F_r \times 4.4}{255 \times A_r}, \quad (10)$$

where $F_r$ is the mean force reading obtained from the sensor during the experiments. It should be noted that $\sigma_0$ is the axial component of $\boldsymbol{\sigma^0}$ in eq. (3) and other two components (lateral and elevation) of $\boldsymbol{\sigma^0}$ are zero.

*D. Calculation of surface area and solidity of the tumors*

The surface area of the tumor, $A_s$ is calculated in cm$^2$ as $A_s = \frac{n_p \times 16}{n_t}$, where $n_p$ is the pixel number inside cancer tumor and $n_t$ is the total number of pixels in the elastogram.

The solidity of the tumor is calculated as [50] $s_t = \frac{a}{c_a}$, where $a$ is the area and $c_a$ is the convex area of the tumor.

*E. Statistical Analysis*

Data in Figs. 5 and 6 are presented as mean $\pm$ SD (standard deviation). Statistical significance was determined using the Kruskal-Wallis test. Matlab (MathWorks Inc., Natick, MA, USA) was used to analyze the data.

## V. Computational considerations

*A. Estimation of axial and lateral displacements and strains*

To compute the axial and lateral strains in both the simulated and experimental data, a technique recently developed in our lab [16] was used.

*B. Estimation of YM and PR*

The axisymmetric assumption in eq. (8) does not produce significant error ($< 6\%$) in inclusions with different shapes, including ellipsoidal inclusions having all three semi-axes of very different length as proven in supplementary information (section 19-A). For this reason, eq. (8) was used to reconstruct the YM and PR in all experimental cases considered in this study (both in vitro and in vivo). For the same reason, only axisymmetric samples were used for the FE and ultrasound simulation investigations. For the FE and ultrasound simulation data, the background strains are computed in a square region of $5 \times 5$ pixels in the left corner of the axial and lateral strain elastograms ($128 \times 128$ pixels). The mean strains of this area are assumed to be representative of the axial and lateral strains of the background region. For the experiments, the background strains are computed in a square region of $10 \times 10$ pixels in the normal tissue away from the tumor or any other organ or tissue abnormality. For the estimation of the YM and PR using the proposed method, non linear least square optimization by 'trust-region-reflective' algorithm in MATLAB (The MathWorks, Natick, MA) is used to minimize the cost function $J$ in eq. 8, where the maximum number of iteration is set to $100$. Complex shapes such as tetragon, pentagon and hexagon are approximated with ellipses, and the cost function for the elliptical tumor has been used for these shapes. The approximation of these complex shapes with ellipses is shown in Fig. S19. The lower and higher limits for the YM are set to $0.1 \times \frac{\sigma_0}{\epsilon_{yy}}$ and $100 \times \frac{\sigma_0}{\epsilon_{yy}}$ inside the tumor. The lower limit of PR in the cost function minimization process is set to $-0.8 \times \frac{\epsilon_{xx}}{\epsilon_{yy}}$ and the higher limit is set to $0.495$. Here, $\epsilon_{yy}$ and $\epsilon_{xx}$ are the axial and lateral strains, respectively.

*C. Implementation of the competing methods*

The YM distributions in samples with different mechanical properties reconstructed using the proposed method were compared with the results obtained using two other 3D reconstruction methods, which are referred to as "3DB" [36] and "3DS" [15]. For the YM reconstruction using the 3DB method, the method described in Bilgen et al. [36] is used while for the YM reconstruction using the 3DS method, the method described in Shin et al. [15] is used. When computing the YM by 3DB, the PR is assumed to be $0.495$ in both the inclusion and background. The correctness of the implementations of these methods is verified by matching the results obtained by our implementations with the results reported in their papers for the same simulation conditions.

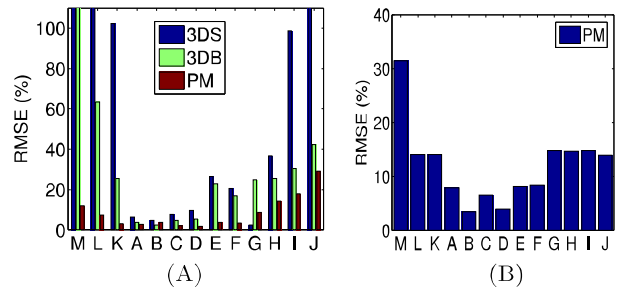

Fig. 1: (A) Percent root mean squared errors (RMSE) of estimated YM images and (B) RMSEs of estimated PR images from three different methods for samples A-M using ultrasound simulated data. RMSEs greater than 100% have been masked to 100%. PM stands for: proposed method. Samples A-J are with tumors harder than the surrounding normal tissue, and K-M are with tumors softer than the surrounding normal tissue. RMSE is higher in case of samples with soft tumors. The RMSEs for 3DB and 3DS method are less than $5\%$ for Sample C, whereas the RMSEs are more than $60\%$ for sample L. RMSE for the proposed method is below $3\%$ for sample C and below $6\%$ for sample L. Sample C and sample L have $5$ and $\frac{1}{5}$ contrast of YM between the tumor and normal tissue. For hard tumors, RMSE in estimating the YM for all three methods increases as the contrast of YM between the tumor and normal tissue increases. The RMSE in estimating the YM by the proposed method is the lowest in all cases in comparison to the other two methods.

## VI. Results

*A. Simulations*

Fig. S3 shows the YM images of the FE simulated samples A-D (see Table S1) reconstructed using the three reconstruction approaches. The corresponding PR images created using our proposed method are shown in Fig. S4. We have also validated local stress measurements obtained using the proposed technique with FE analysis and demonstrated excellent agreement between theory and simulations ($< 1\%$ error) in samples A-D (see supplementary information (section 22)).

The RMSE occurring when reconstructing the YM distribution in tumors of different shapes from FE simulations using the 3 approaches are shown in Table I. We observe that, within the results obtained using the proposed approach, the highest RMSE is observed when the shape of the tumor is cylindrical ($9.91\%$) and the lowest ($0.8\%$) when the shape of the tumor is spherical. In comparison, the RMSE associated to the other two YM reconstruction approaches are much higher than the one associated to the proposed method and typically higher than 20% for tumors of all shapes. The RMSEs occurring when reconstructing the PRs using the proposed approach are also shown in the table, and they are found to be less than 9% in all simulated samples.

The RMSEs computed for the three methods in the case of tumors having different YM contrast (CTYM) with respect to the background are reported in Table II. We see that the RMSE associated to the proposed approach is below 5% for contrast of $0.1 - 100$, whereas the RMSEs for the 3DB and 3DS approaches are higher than 20% in most cases.

A typical problem of elastography-based reconstruction methods is the effect of boundary conditions on the reconstructed mechanical parameters. The RMSEs computed when the YM of the tumor is reconstructed using data obtained with different boundary conditions are shown in Table III. We see that, even in the case of very complex boundary conditions, the proposed approach can reconstruct the YM with about 90% accuracy. The other two reconstruction methods, instead, show larger RMSEs for all boundary conditions when compared to the proposed one.

TABLE I: RMSEs in estimating the YM of inclusions of different shapes using different methods when the YM inclusion/background contrast (CTYM) is 3. RMSEs in estimating the PR in the same inclusions using the proposed method (PM) are shown in parentheses.

| Sample | Shape | 3DB (%) | 3DS (%) | PM (%) |
|---|---|---|---|---|
| Z1 | spherical | 16.95 | 2.38 | 0.8(0.5) |
| Z2 | prolate | 26.86 | 3.28 | 2.69 (2.31) |
| Z3 | oblate | 56.00 | 27.06 | 5.01 (2.66) |
| Z4 | cylindrical | 21.57 | 22.18 | 9.91 (8.88) |
| Z5 | penny | 62.95 | 129.73 | 8.29 (12.67) |
| Z6 | 3D tetragon | 27.39 | 43.00 | 9.39 (7.04) |
| Z7 | 3D pentagon | 21.34 | 24.75 | 8.55(7.31) |
| Z8 | 3D hexagon | 41.73 | 46.42 | 8.12(6.75) |

TABLE II: RMSEs in estimates of YM of spherical shaped inclusions by different methods for different CTYM values. RMSEs in estimating the PR in the same inclusions using the proposed method (PM) are shown in parentheses.

| Sample | CTYM | 3DB (%) | 3DS (%) | PM (%) |
|---|---|---|---|---|
| X1 | 0.1 | 888.97 | 885.69 | 3.54(0.34) |
| X2 | 0.2 | 47.59 | 393.18 | 2.25 (0.19) |
| X3 | 0.5 | 4.54 | 97.71 | 1.34 (0.08) |
| X4 | 3 | 22.77 | 26.60 | 0.56 (0.05) |
| X5 | 5 | 23.77 | 26.35 | 0.64 (0.39) |
| X6 | 15 | 24.58 | 2.52 | 1.71 (0.94) |
| X7 | 25 | 24.53 | 36.54 | 2.34(3.36) |
| X8 | 50 | 25.15 | 98.72 | 3.85(11.98) |
| X9 | 100 | 24.59 | 115.81 | 4.75(2.41) |

Effect of heterogeneity in the YM distribution inside the tumor on the reconstructed parameters has been investigated, and the results are reported in Table IV. Sample H3 has the highest heterogeneity, where YM reduces by 30% from the center to the periphery of the tumor. The proposed method is capable of reconstructing the YM of the tumor with high accuracy ($> 94\%$) in all cases analyzed in this study whereas the other two approaches introduce more than 14% error in most cases.

TABLE III: RMSEs in estimating the YM of spherically shaped inclusions with a CTYM of 3 by using different methods for different complex boundary conditions. RMSEs in estimating the PR in the same inclusions using the proposed method (PM) are shown in parentheses.

| Sample | Boundary condition | 3DB (%) | 3DS (%) | PM (%) |
|---|---|---|---|---|
| B1 | zig-zag stiffer tissue in background | 19.25 | 21.82 | 7.74(8.32) |
| B2 | 14 different shaped inclusions in background | 30.90 | 34.91 | 10.94 (3.67) |
| B3 | strip of hard tissue on top of the tumor | 16.56 | 19.94 | 0.9 (2.14) |
| B4 | multiple layers of tissue on top of the tumor | 12.03 | 14.87 | 5.06 (5.02) |
| B5 | multiple layers of tissue on top of the tumor | 21.86 | 24.15 | 5.55 (6.37) |

TABLE IV: RMSEs in estimating the YM of the spherical shaped inclusions of CTYM of 3 by different methods for different heterogeneity conditions. RMSEs in estimating the PR in the same inclusions using the proposed method (PM) are shown in parentheses.

| Sample | Heterogeneity (%) | 3DB (%) | 3DS (%) | PM (%) |
|---|---|---|---|---|
| H1 | 10 | 16.12 | 18.98 | 2.27(4.13) |
| H2 | 20 | 14.79 | 16.78 | 3.91 (5.25) |
| H3 | 30 | 13.95 | 14.91 | 6.22 (11.45) |

The results related to the non-uniform compression conditions from FE simulations are shown in Table V. Once again, the proposed method is robust to load variations, as opposed to the other two methods.

In Fig. 1(A), we report the RMSEs of the estimated YM images using the three reconstruction techniques for thirteen samples A-M when using simulated ultrasound data, and the RMSEs of the estimated PRs in the same samples using the proposed technique are shown in Fig. 1(B). The reconstructed YM images of sample A-D are shown in Fig. S11 and the reconstructed PRs of samples A-D are shown in Fig. S12. We see from Fig. 1 that the error incurred in all the reconstruction methods increases as the inclusion/background YM contrast increases. However, the RMSE for the proposed method is below or around 15% for inclusion/background YM contrasts up to 50 (sample H). The RMSE for the estimated PR also increases as the inclusion/background YM contrast increases but remains around 10% even in case of a YM contrast of 100 (sample J). The other two methods can introduce errors greater than 25% even in case of a contrast as low as 3 (sample E). The RMSE for all methods increases for the samples with a soft inclusion (samples K-M). However, the error is significantly lower for the proposed method in comparison to other two techniques. For sample M, where the inclusion is 10 times softer than the background, the RMSEs are higher than 100% for 3DB and 3DS methods, while the RMSE for the proposed technique is below 10%. These results prove that the proposed approach is more accurate, more precise and more robust than previously proposed 3D YM reconstruction methods and has the advantage to provide estimates of both the YM and the PR distributions.

TABLE V: RMSEs in estimating the YM of spherical shaped inclusions with CTYM of 3 by different methods under non-uniform loading. RMSEs in estimating the PR in the same inclusions using the proposed method (PM) are shown in parentheses. In these cases, the load is increased or decreased from the center to the periphery of the compressor plate.

| Sample | Non-uniformity of loading (%) | 3DB (%) | 3DS (%) | PM (%) |
|---|---|---|---|---|
| R1 | 20% reduction | 31.84 | 35.54 | 9.47(1.89) |
| R2 | 10% reduction | 24.68 | 28.26 | 5.24 (1.02) |
| R3 | 20% increment | 7.94 | 10.86 | 9.52 (3.20) |
| R4 | 10% increment | 12.96 | 15.94 | 6.45 (3.32) |

*B. Controlled experiments*

Fig. 2 shows selected results from controlled experiments performed on a breast-mimicking phantom containing different spherical inclusions simulating incompressible tumors with similar stiffness. In Fig. 2, the estimated axial strain, lateral strain, reconstructed YM and PR distributions for one of the inclusions inside the breast phantom are shown. The mean and standard deviation values of the reconstructed YM and PR distributions as obtained from this experiment can be found in Table VI. According to the manufacturer's specifications for this breast phantom [43]–[45], the YM of the background is $20 \pm 5$ kPa while the YM in the inclusion is approximately $50$ kPa. The PR is approximately $0.5$ both in the inclusion and in the background [46]. Thus, our reconstructed YM and PR have errors less than $7\%$ and $15\%$, respectively.

In Table S9, we show the mean values of the reconstructed YM and PR obtained from four additional experiments performed on the breast phantom. From these data, we observe that the mean applied force is different in all four controlled experiments. However, the mean values of the reconstructed YM and PR in the inclusion and background are very similar with errors of $< 10\%$ and $< 16\%$, respectively. These results prove the robustness of the proposed method to variations in the applied load. The reconstructed YM in the breast phantom was also independently validated using a shear wave elastography system [51], [52] as reported in supplementary information (section 21).

*C. In vivo experiments*

B-mode images and reconstructed YM and PR distributions obtained from data acquired from three untreated mice at three different time points (week 1, week 2 and week 3) are shown in Fig. 3 (A1-A9, B1-B9 and C1-C9). We see from this figure that, in general, the YM increases significantly from week 1 (A2, A5, A8) to week 3 (C2, C5, C8) in the untreated mice, while the PR values are around $0.25$ to $0.35$. Based on prior literature on elastography, we expect most cancers to be stiffer than the normal tissue. However, to the best of our knowledge, our results are the first ones to experimentally demonstrate the actual increase of YM as the cancer progresses using ultrasound compressive elastography.

B-mode images and reconstructed YM and PR distributions obtained from data acquired from three treated mice at three different time points (week 1, week 2 and week 3) are shown in Fig. 4 (A1-A9, B1-B9 and C1-C9). We see from this figure that, in most treated mice, the YM decreases or does not change with time. Also, the YM contrast between cancer and background tissue is not as high as in the case of the untreated mice. The PR values are in the range $0.3 - 0.4$ in most of the cases. However, the PR appears to increase in the first or second week and then to decrease in the third week in most of the cases. Once again, to our knowledge, these YM and PR trends in tumors following a treatment have not been experimentally investigated using compressive elastography prior to this study.

The YM mean values with the corresponding standard deviations for twelve treated mice and seven untreated mice at the three different time points (week 1, week 2 and week 3) are shown in Fig. 5 (A1). In the first week, the mean YM in the untreated tumors was found to be below $50$ kPa. In the second week, the mean YM in the untreated tumors increased significantly ( above $60$ kPa) and in the third week was found to be above $75$ kPa. The mean YM in the treated tumors at the three different weeks was found to be close to $25$ kPa, which is a value close to the one measured for the YM in the normal tissue (background).

The mean values of PR with the corresponding standard deviations for all the treated and untreated mice at the three different time points (week 1, week 2 and week 3) are shown in Fig. 5 (A2). In all these time points, the treated mice were found to have higher PR than the untreated ones. For both the treated and untreated mice, the mean PR does not appear to change significantly at the different time points.

Fig. 5 (A3) shows the tumor/background YM contrast for the twelve treated and seven untreated cancers, while Fig. 5 (A4) shows the tumor/background PR contrast (CTPR) for the twelve treated and seven untreated cancers. In Fig. 5 (A3), we see that the CTYM for untreated cancers is higher than that for the treated ones in all three weeks, which confirms previously reported findings [53].

Mean surface areas of the tumors with the corresponding standard deviations for all the treated and untreated mice at the three time points are shown in Fig. 6 (A1). The mean surface area of the treated tumors does not change significantly with time, whereas the mean surface area of the untreated tumors increases with time.

The solidity of the tumor is a measure of the regularity of the shape of the tumor, and the mean values of solidity for all the tumors at the three time points are shown in Fig. 6 (A2). Solidity is higher in the case of the treated tumors than in the case of the untreated tumors at all time points. In a previous study, lower values of solidity have been associated to malignancy [54].

VII. DISCUSSION

In this paper, we propose a new, non-invasive, three-dimensional method for simultaneously reconstructing both the YM and PR in tumors. The YM is a mechanical parameter that has been investigated as a marker for cancer diagnosis, prognosis and treatment monitoring and planning. PR is another mechanical parameter, whose role in cancer assessment has not been fully elucidated yet, but it has been shown to have potentials in cancer-related diseases such as lymphedema [17], [18]. Both these parameters are needed to estimate other important properties of a tumor such as interstitial permeability and vascular permeability [27].

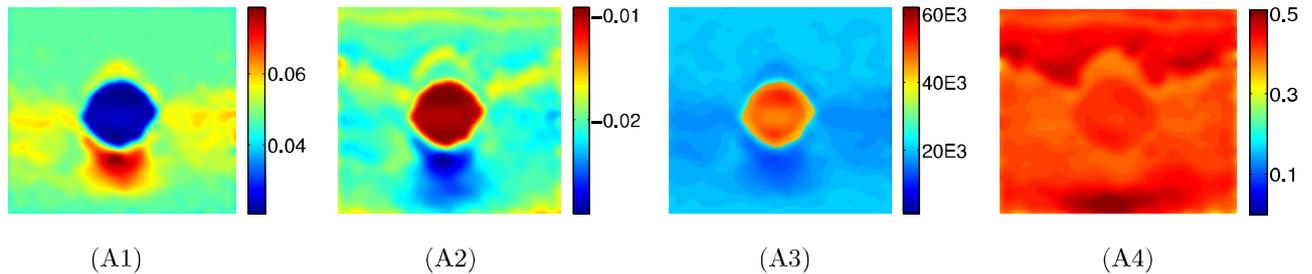

(A1) (A2) (A3) (A4)

Fig. 2: Estimated axial strain (A1), lateral strain (A2), YM (in Pa) image (A3) and PR image (A4) from the controlled experiment (CE1). This figure shows results for applied compression of $0.97$ kPa. The estimated YM is in the range of $45-55$ kPa in the inclusion and in the range of $17-21$ kPa in the background region. The estimated PR is around $0.44$ in the inclusion region and $0.43$ in the background region.

TABLE VI: Mean and standard deviation of the reconstructed YM and PR distributions in controlled experiment

| Exp no | Applied load (kPa) | Est YM of inclusion (kPa) | Est PR of inclusion | Est YM of background (kPa) | Est PR of background |
|---|---|---|---|---|---|
| CE1 | 0.97 | 46.72 ±3.59 | 0.432 ±0.02 | 19.27 ±1.61 | 0.427 ±0.03 |

The proposed method has many advantages compared to previously proposed reconstruction techniques, which are currently used in elastography. It can accurately reconstruct the YM and PR of a tumor for a wide range of tumor/background YM contrast, in many complex boundary conditions and independently of the shape of the tumor. The proposed method is also robust to practical experimental conditions that may deviate from the ideal ones such as non-uniform loading and when the YM inside the tumor is heterogeneous. Thus, the proposed method has the potential to significantly improve the way the YM of tumors is currently imaged and quantified as well as to provide a new means to image and quantify the PR of tumors and normal tissues in vivo.

Based on our in vivo animal results, YM in the untreated tumors was found to be increasing with time, whereas the YM in the treated ones did not change significantly with time. In most of the cancers (both treated and untreated), we found out that the PR is higher in the tumor than in the soft tissue. The values of PR found in this study match well with those previously reported in the literature as estimated using invasive techniques [26], [29]–[31]. In our in vivo experiments, we have obtained PR values in the range $0.25-0.35$ in the tumors and in the range of $0.2-0.35$ in the normal tissue. Based on the model proposed in Boccaccini et al. [55], PR values in the range $0.2-0.35$ correspond to underlying porosity values in the range $0.13-0.32$, which would be consistent with previously reported values of porosity in tumors and normal tissue [56], [57]. The shape regularity index (solidity) and surface area of the tumors were also used to further characterize the in vivo results.

It is a common assumption in many studies reported in the literature pertaining elastography to treat tumors and soft tissues as incompressible elastic solids [10], [12]–[15]. Our study shows that, if not fully satisfied, such assumption can lead to significant errors in the reconstructed YM values (Table II sample X4) even in the case of small YM tumor/background contrasts, and that this error increases as the YM tumor/background contrast increases. Thus, accurate estimation of the PR is not only important because of its potential to provide new clinical information but also to obtain accurate estimates of the YM distribution.

Our proposed method is a three-dimensional method, where, ideally, the three normal components of the strain tensor (axial, lateral and elevational) are used to reconstruct the YM and PR in the tissue (see eq. 6). However, experimental estimation of the elevational strain component requires data acquisition along the elevational direction [13], [58], [59], which is difficult to accomplish in free-hand elastography, in general, and particularly challenging in poroelastography, where temporal data is required. Currently, most clinical studies involving ultrasound imaging use linear arrays, which allow only planar data acquisition similarly to what we have reported in this manuscript. Therefore, implicit to this and most of the reconstruction procedures proposed in elastography is the assumption that the normal lateral and elevational components of the strain are identical. We have proved with extensive simulations and error analysis that such assumption introduces small errors ($< 6\%$) in the YM and PR reconstructed using our proposed method. This has been demonstrated in a number of three-dimensional samples with different geometry, covering most cases of practical interest for clinical elastography (see supplementary information (section 19-A)).

Plane strain and plane stress conditions have been assumed in a number of previously proposed YM reconstruction methods [22], [40]. Although these assumptions are typically not valid in applications of clinical interest, we show in supplementary information (section 19-B) that our method can reconstruct the YM and PR with high accuracy even in special cases where these conditions may be assumed as long as the three normal strain components are available.

Although the YM and PR reconstruction formulations were derived for a sample with a single inclusion, we demonstrate that the proposed theory can also be used to accurately reconstruct the YM and PR in samples containing a number of inclusions, by reconstructing the YM and PR of each of the inclusions separately (see supplementary information (section 19-C)).

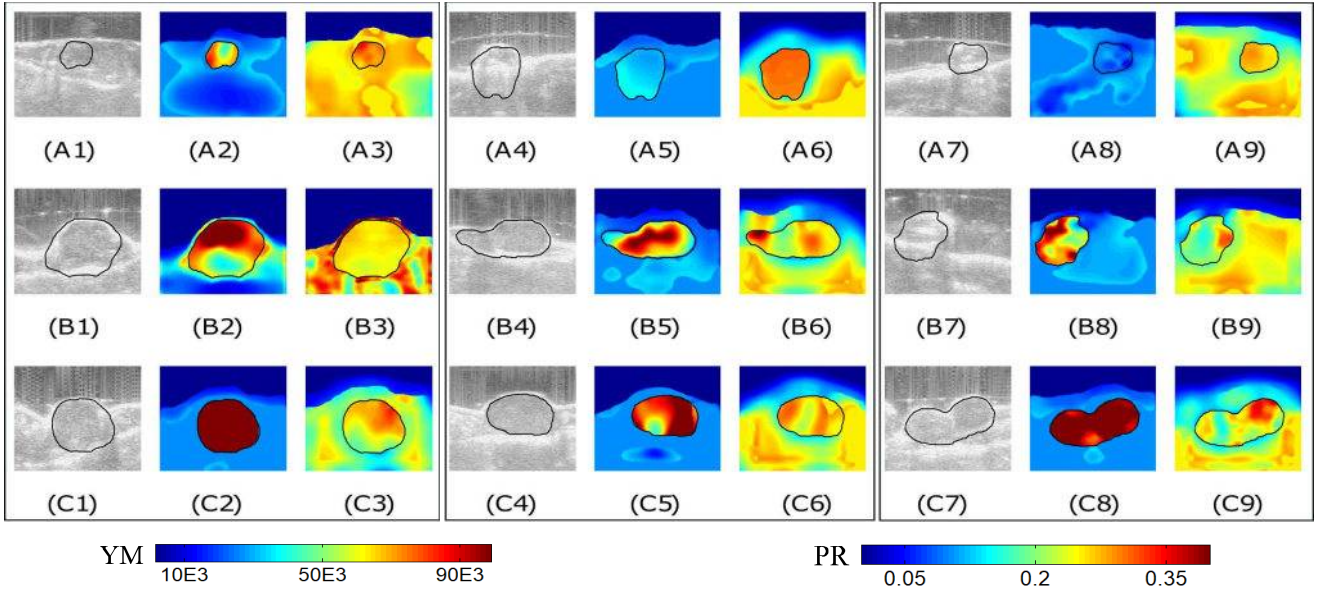

Fig. 3: B-mode images of untreated mouse #1 at three time points (week 1, week 2, week 3) are shown in (A1), (B1) and (C1). Reconstructed YM (in Pa) and PR distributions at the same time points are shown in (A2), (B2) and (C2), and (A3), (B3) and (C3), respectively. B-mode images of untreated mouse #2 at three time points (week 1, week 2, week 3) are shown in (A4), (B4) and (C4). Reconstructed YM (in Pa) and PR distributions at the same time points are shown in (A5), (B5) and (C5), and (A6), (B6) and (C6), respectively. B-mode images of untreated mouse #3 at three time points (week 1, week 2, week 3) are shown in (A7), (B7) and (C7). YM (in Pa) and PR distributions at the same time points are shown in (A8), (B8) and (C8), and (A9), (B9) and (C9), respectively. The YMs for the three cases increase from week 1 to week 3. More specifically, the YMs for the shown untreated mice are below $50$ kPa in the first week, around $80$ kPa in the second week and more than $90$ kPa in the third week. These results indicate the increasing hardening of the tumor as the cancer progresses. The PRs do not change significantly at the three time points ($\approx 0.3$).

In this study, we have compared the performance of our method with two other techniques, which are three dimensional and do not require assumptions on the boundary conditions at the side borders of the sample. Among these two techniques, the first one is a recent one [15] and the other one is a classical technique [36]. Other reconstruction methods have been proposed in the literature [58], [59], which are also three-dimensional and may have some advantages with respect to the methods in Refs. [15] [36] chosen for the statistical comparison carried out in this paper. However, these methods require assumptions about the boundary conditions and about the incompressibility of the tissue, and, in some cases, very controlled experimental data acquisition protocols, which are of difficult implementation in current poroelastography experiments.

There are several factors that can affect the accuracy of the reconstructed YM and PR values using the proposed approach. The first one is the quality of the axial strain and lateral strain estimates. The proposed method is able to reconstruct YM and PR with an error of or below $5\%$ for a tumor/background YM contrast of $0.1 - 100$, when the estimations of the axial and lateral strains are error- and noise-free such as those directly obtained from FE simulations. However, it is known that lateral strain estimation in elastography is typically noisier than axial strain estimation. We have recently proposed a new method capable of providing high quality lateral strain estimations [16]. This method has been used in the study reported in this paper. Another important factor affecting reconstruction is the YM contrast between the tumor and the background. The axial strain ratio and lateral strain ratio between the tumor and normal tissue saturate for very small or large YM contrasts (discussed in supplementary information (section 7)), which is a fundamental limitation also referred to as contrast transfer efficiency in elastographic problems dealing with non-uniform materials [15], [40]. Because of this fundamental limitation, all elastographic reconstruction algorithms including the proposed one fail to accurately determine the YM and PR when the YM contrast is very small or large. In this paper, we demonstrate that our method can estimate the YM and PR with moderate accuracy (error around $15\%$) for YM tumor/background contrasts in the range of $0.1 - 50$, even in the presence of ultrasound noise (Fig. 1 (A)). We believe that this range should cover practical cancer imaging scenarios [60] and denotes a far superior performance with respect to previously reported reconstruction methods. Other challenges in the accurate estimation of YM and PR from practical elastography experiments are the target hardening effect, non-uniformity of applied stress and decorrelation noise due to the out of plane and other uncontrollable motion. We have discussed these issues in supplementary information (sections 8, 25 and 26, respectively). Requirement of high computational effort is a limitation of the proposed method, which we have discussed in supplementary information (section 27).

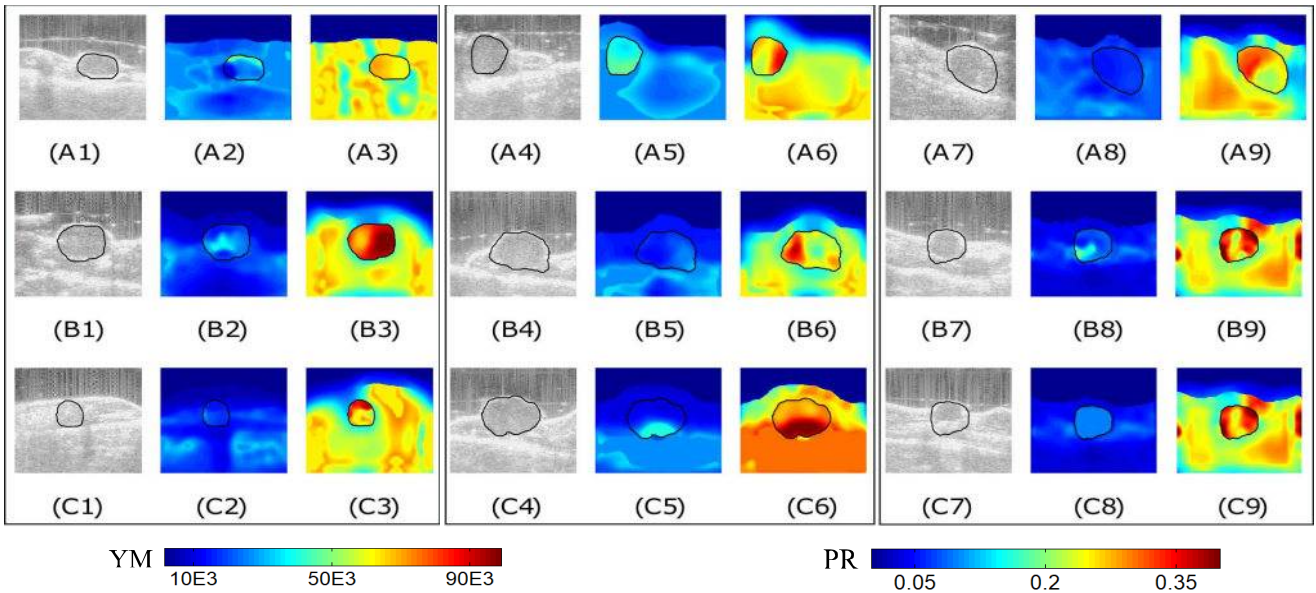

Fig. 4: B-mode images of treated mouse $\#1$ at three time points (week 1, week 2, week 3) are shown in (A1), (B1) and (C1). Reconstructed YM (in Pa) and PR distributions at the same time points are shown in (A2), (B2) and (C2), and (A3), (B3) and (C3), respectively. B-mode images of treated mouse $\#2$ at three time points (week 1, week 2, week 3) are shown in (A4), (B4) and (C4). Reconstructed YM (in Pa) and PR distributions at the same time points are shown in (A5), (B5) and (C5), and (A6), (B6) and (C6), respectively. B-mode images of treated mouse $\#3$ at three time points (week 1, week 2, week 3) are shown in (A7), (B7) and (C7). Reconstructed YM (in Pa) and PR distributions at the same time points are shown in (A8), (B8) and (C8), and (A9), (B9) and (C9), respectively. The YMs for these treated mice are around $20$ kPa for all time points, and the PRs are around $0.35$ for all time points. Overall, the YM values of the treated mice are significantly lower than that of the untreated mice, whereas the PR values of the treated mice are higher than that of the untreated ones. The reduction/non-increment of stiffness of the treated tumors may be a sign of the efficacy of the treatment in controlling the growth of the cancer.

## VIII. Conclusions

In this paper, we have developed a three dimensional reconstruction method based on Eshelby's formulation for materials with inclusion. Our proposed method can accurately estimate and image both the PR and YM of tumors and surrounding tissue and is robust to changes of complex boundary conditions of the tumor environment and the shape of the tumor. Simulations and controlled ultrasound elastography experiments unequivocally demonstrate that the proposed method is capable of reconstructing these parameters accurately in many experimental scenarios of clinical relevance. Based on the potential role of YM and PR as markers for cancer diagnosis, prognosis and treatment efficacy, the proposed method can have a significant impact in the assessment of cancers and, in general, in the field of cancer elasticity imaging.

## References

[1] L. Sandrin, B. Fourquet, J.-M. Hasquenoph, S. Yon, C. Fournier, F. Mal, C. Christidis, M. Ziol, B. Poulet, F. Kazemi et al., "Transient elastography: a new noninvasive method for assessment of hepatic fibrosis," Ultrasound in medicine & biology, vol. 29, no. 12, pp. 1705–1713, 2003.

[2] C. L. De Korte, G. Pasterkamp, A. F. Van Der Steen, H. A. Woutman, and N. Bom, "Characterization of plaque components with intravascular ultrasound elastography in human femoral and coronary arteries in vitro," Circulation, vol. 102, no. 6, pp. 617–623, 2000.

[3] J. D. Eshelby, "The determination of the elastic field of an ellipsoidal inclusion, and related problems," in Proceedings of the Royal Society of London A: Mathematical, Physical and Engineering Sciences, vol. 241, no. 1226. The Royal Society, 1957, pp. 376–396.

[4] H. T. Nia, H. Liu, G. Seano, M. Datta, D. Jones, N. Rahbari, J. Incio, V. P. Chauhan, K. Jung, J. D. Martin et al., "Solid stress and elastic energy as measures of tumour mechanopathology," Nature Biomedical Engineering, vol. 1, p. 0004, 2016.

[5] P. A. Netti, L. T. Baxter, Y. Boucher, R. Skalak, and R. K. Jain, "Macro- and microscopic fluid transport in living tissues: Application to solid tumors," AIChE journal, vol. 43, no. 3, pp. 818–834, 1997.

[6] P. A. Netti, D. A. Berk, M. A. Swartz, A. J. Grodzinsky, and R. K. Jain, "Role of extracellular matrix assembly in interstitial transport in solid tumors," Cancer research, vol. 60, no. 9, pp. 2497–2503, 2000.

[7] J. Ophir, S. Alam, B. Garra, F. Kallel, E. Konofagou, T. Krouskop, and T. Varghese, "Elastography: ultrasonic estimation and imaging of the elastic properties of tissues," Proceedings of the Institution of Mechanical Engineers, Part H: Journal of Engineering in Medicine, vol. 213, no. 3, pp. 203–233, 1999.

[8] F. Sebag, J. Vaillant-Lombard, J. Berbis, V. Griset, J. Henry, P. Petit, and C. Oliver, "Shear wave elastography: a new ultrasound imaging mode for the differential diagnosis of benign and malignant thyroid nodules," The Journal of Clinical Endocrinology & Metabolism, vol. 95, no. 12, pp. 5281–5288, 2010.

[9] R. Muthupillai and R. L. Ehman, "Magnetic resonance elastography," Nature medicine, vol. 2, no. 5, pp. 601–603, 1996.

[10] F. Kallel and M. Bertrand, "Tissue elasticity reconstruction using linear perturbation method," IEEE Transactions on Medical Imaging, vol. 15, no. 3, pp. 299–313, 1996.

[11] M. Doyley, P. Meaney, and J. Bamber, "Evaluation of an iterative reconstruction method for quantitative elastography," Physics in medicine

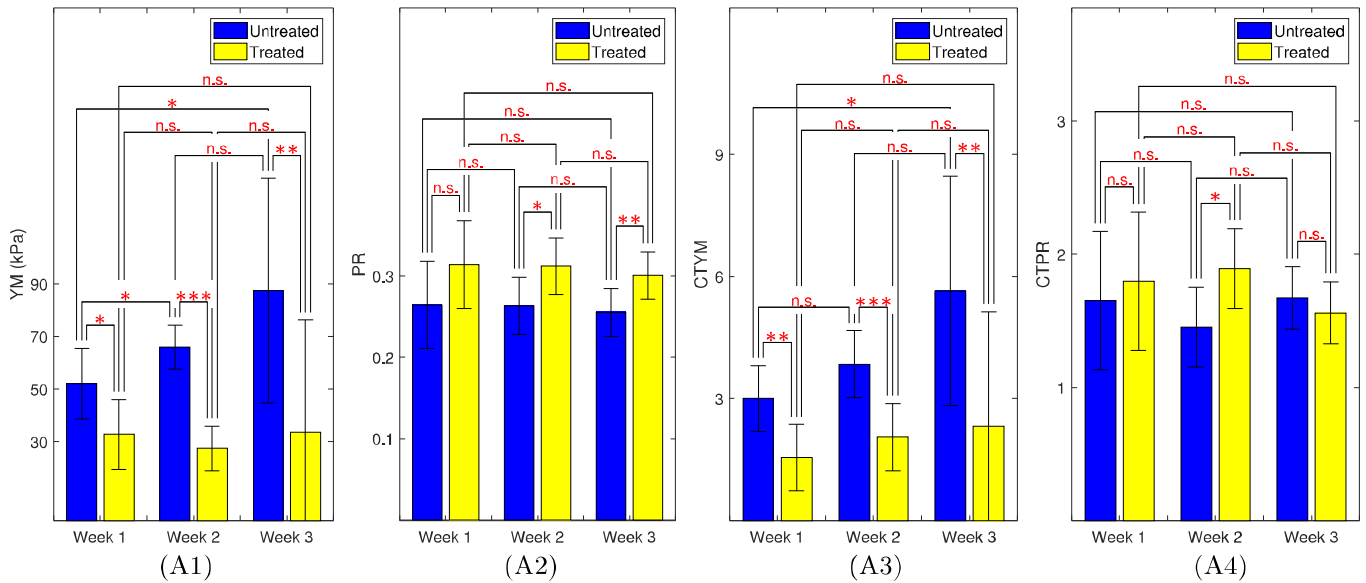

Fig. 5: (A1) Mean YM values for the treated and untreated mice at week 1, week 2 and week 3. (A2) Mean PR values for the treated and untreated mice at week 1, week 2 and week 3. (A3) Mean YM contrast between tumor and normal tissue for treated and untreated mice at week 1, week 2 and week 3. (A4) Mean PR contrast between tumor and normal tissue for treated and untreated mice at week 1, week 2 and week 3. n.s. means not statistically significant. One, two and three stars correspond to $p$-value less than $0.05, 0.01, 0.001$, respectively. The mean values of YM and CTYM of the tumors increase from week 1 to week 3 for untreated mice and remain almost constant for the treated ones. Mean values of PR and CTPR of the tumors are consistently higher for the treated tumors than the untreated ones.

*and biology*, vol. 45, no. 6, p. 1521, 2000.

[12] M. M. Doyley, S. Srinivasan, S. A. Pendergrass, Z. Wu, and J. Ophir, "Comparative evaluation of strain-based and model-based modulus elastography," *Ultrasound in medicine & biology*, vol. 31, no. 6, pp. 787–802, 2005.

[13] A. A. Oberai, N. H. Gokhale, M. M. Doyley, and J. C. Bamber, "Evaluation of the adjoint equation based algorithm for elasticity imaging," *Physics in Medicine and Biology*, vol. 49, no. 13, p. 2955, 2004.

[14] J. Fehrenbach, M. Masmoudi, R. Souchon, and P. Trompette, "Detection of small inclusions by elastography," *Inverse problems*, vol. 22, no. 3, p. 1055, 2006.

[15] B. Shin, D. Gopaul, S. Fienberg, and H. J. Kwon, "Application of eshelbys solution to elastography for diagnosis of breast cancer," *Ultrasonic imaging*, vol. 38, no. 2, pp. 115–136, 2016.

[16] M. T. Islam, A. Chaudhry, S. Tang, E. Tasciotti, and R. Righetti, "A new method for estimating the effective poissons ratio in ultrasound poroelastography," *IEEE Transactions on Medical Imaging*, 2018.

[17] R. Righetti, B. S. Garra, L. M. Mobbs, C. M. Kraemer-Chant, J. Ophir, and T. A. Krouskop, "The feasibility of using poroelastographic techniques for distinguishing between normal and lymphedematous tissues in vivo," *Physics in medicine and biology*, vol. 52, no. 21, p. 6525, 2007.

[18] G. P. Berry, J. C. Bamber, P. S. Mortimer, N. L. Bush, N. R. Miller, and P. E. Barbone, "The spatio-temporal strain response of oedematous and nonoedematous tissue to sustained compression in vivo," *Ultrasound in medicine & biology*, vol. 34, no. 4, pp. 617–629, 2008.

[19] E. E. Konofagou, T. P. Harrigan, J. Ophir, and T. A. Krouskop, "Poroelastography: imaging the poroelastic properties of tissues," *Ultrasound in medicine & biology*, vol. 27, no. 10, pp. 1387–1397, 2001.

[20] K. Raghavan and A. E. Yagle, "Forward and inverse problems in elasticity imaging of soft tissues," *IEEE Transactions on nuclear science*, vol. 41, no. 4, pp. 1639–1648, 1994.

[21] A. Skovoroda, S. Emelianov, and M. O'donnell, "Tissue elasticity reconstruction based on ultrasonic displacement and strain images," *IEEE transactions on ultrasonics, ferroelectrics, and frequency control*, vol. 42, no. 4, pp. 747–765, 1995.

[22] C. Sumi, A. Suzuki, and K. Nakayama, "Estimation of shear modulus distribution in soft tissue from strain distribution," *IEEE Transactions on Biomedical Engineering*, vol. 42, no. 2, pp. 193–202, 1995.

[23] A. A. Oberai, N. H. Gokhale, and G. R. Feijóo, "Solution of inverse problems in elasticity imaging using the adjoint method," *Inverse problems*, vol. 19, no. 2, p. 297, 2003.

[24] A. Samani, J. Bishop, and D. B. Plewes, "A constrained modulus reconstruction technique for breast cancer assessment," *IEEE Transactions on Medical Imaging*, vol. 20, no. 9, pp. 877–885, 2001.

[25] J. Fehrenbach, "Influence of poisson's ratio on elastographic direct and inverse problems," *Physics in medicine and biology*, vol. 52, no. 3, p. 707, 2007.

[26] T. Stylianopoulos, J. D. Martin, M. Snuderl, F. Mpekris, S. R. Jain, and R. K. Jain, "Coevolution of solid stress and interstitial fluid pressure in tumors during progression: implications for vascular collapse," *Cancer research*, vol. 73, no. 13, pp. 3833–3841, 2013.

[27] R. Leiderman, P. E. Barbone, A. A. Oberai, and J. C. Bamber, "Coupling between elastic strain and interstitial fluid flow: ramifications for poroelastic imaging," *Physics in medicine and biology*, vol. 51, no. 24, p. 6291, 2006.

[28] M. Bayat, A. Nabavizadeh, V. Kumar, A. Gregory, M. Insana, A. Alizad, and M. Fatemi, "Automated in vivo sub-hertz analysis of viscoelasticity (save) for evaluation of breast lesions," *IEEE Transactions on Biomedical Engineering*, 2017.

[29] F. Mpekris, J. W. Baish, T. Stylianopoulos, and R. K. Jain, "Role of vascular normalization in benefit from metronomic chemotherapy," *Proceedings of the National Academy of Sciences*, vol. 114, no. 8, pp. 1994–1999, 2017.

[30] Y.-C. Fung, "Mechanical properties and active remodeling of blood vessels," in *Biomechanics*. Springer, 1993, pp. 321–391.

[31] T. Roose, P. A. Netti, L. L. Munn, Y. Boucher, and R. K. Jain, "Solid stress generated by spheroid growth estimated using a linear poroelasticity model," *Microvascular research*, vol. 66, no. 3, pp. 204–212, 2003.

[32] N. I. Nikolaev, T. Müller, D. J. Williams, and Y. Liu, "Changes in the stiffness of human mesenchymal stem cells with the progress of cell death as measured by atomic force microscopy," *Journal of biomechanics*, vol. 47, no. 3, pp. 625–630, 2014.

[33] D. C. Stewart, A. Rubiano, K. Dyson, and C. S. Simmons, "Mechanical characterization of human brain tumors from patients and comparison

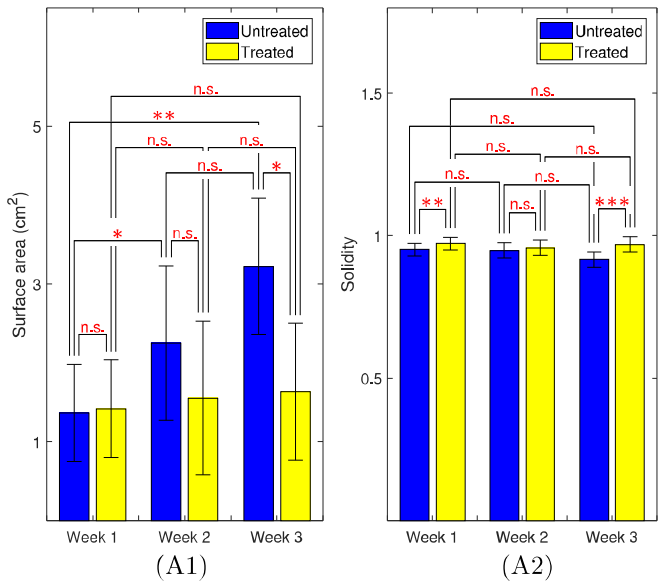

Fig. 6: (A1) Mean surface areas of the tumors for the treated and untreated mice at week 1, week 2 and week 3. (A2) Mean values of solidity for all treated and untreated mice at week 1, week 2 and week 3. n.s. means not statistically significant. One, two and three stars correspond to $p$-value less than $0.05, 0.01, 0.001$, respectively. The mean value of surface area of the tumors increases from week 1 to week 3 for untreated mice and remains almost constant for the treated ones. Mean value of solidity of the tumors is consistently higher for the treated tumors than the untreated ones.

to potential surgical phantoms," *PloS one*, vol. 12, no. 6, p. e0177561, 2017.

[34] D. Thanoon, M. Garbey, N.-H. Kim, and B. Bass, "A computational framework for breast surgery: application to breast conserving therapy," in *Computational Surgery and Dual Training*. Springer, 2010, pp. 249–266.

[35] C. Voutouri, C. Polydorou, P. Papageorgis, V. Gkretsi, and T. Stylianopoulos, "Hyaluronan-derived swelling of solid tumors, the contribution of collagen and cancer cells, and implications for cancer therapy," *Neoplasia*, vol. 18, no. 12, pp. 732–741, 2016.

[36] M. Bilgen and M. F. Insana, "Elastostatics of a spherical inclusion in homogeneous biological media," *Physics in Medicine and Biology*, vol. 43, no. 1, p. 1, 1998.

[37] P. E. Barbone and N. H. Gokhale, "Elastic modulus imaging: on the uniqueness and nonuniqueness of the elastography inverse problem in two dimensions," *Inverse problems*, vol. 20, no. 1, p. 283, 2004.

[38] O. A. Babaniyi, A. A. Oberai, and P. E. Barbone, "Direct error in constitutive equation formulation for plane stress inverse elasticity problem," *Computer methods in applied mechanics and engineering*, vol. 314, pp. 3–18, 2017.

[39] C. Jia, S. Alam, R. Azar, and B. Garra, "Estimation of shear modulus ratio between inclusion and background using strain ratios in 2-d ultrasound elastography," *IEEE Transactions on Ultrasonics, Ferroelectrics, and Frequency Control*, vol. 61, no. 4, pp. 611–619, 2014.

[40] F. Kallel, M. Bertrand, and J. Ophir, "Fundamental limitations on the contrast-transfer efficiency in elastography: an analytic study," *Ultrasound in medicine & biology*, vol. 22, no. 4, pp. 463–470, 1996.

[41] T. Liu, O. A. Babaniyi, T. J. Hall, P. E. Barbone, and A. A. Oberai, "Noninvasive in-vivo quantification of mechanical heterogeneity of invasive breast carcinomas," *PloS one*, vol. 10, no. 7, p. e0130258, 2015.

[42] T. Mura, "Micromechanics of defects in solids. mechanics of elastic and inelastic solids, vol. 3," 1987.

[43] http://www.cirsinc.com/products/all/83/-breast-elastography-phantom/, 2018, [Online; accessed 15-January-2018].

[44] J. Yue, M. Tardieu, F. Julea, L. Chami, O. Lucidarme, X. Maître, and C. Pellot-Barakat, "Comparison between 3d supersonic shear wave elastography and magnetic resonance elastography: a preliminary experimental study," in *Journées RITS 2015*, 2015, pp. pp–142.

[45] S. Cournane, A. Fagan, and J. Browne, "Review of ultrasound elastography quality control and training test phantoms," *Ultrasound*, vol. 20, no. 1, pp. 16–23, 2012.

[46] K. Hollerieth, B. Gaßmann, S. Wagenpfeil, P. Moog, M.-T. Vo-Cong, U. Heemann, and K. F. Stock, "Preclinical evaluation of acoustic radiation force impulse measurements in regions of heterogeneous elasticity," *Ultrasonography*, vol. 35, no. 4, p. 345, 2016.

[47] R. Palomba, A. Parodi, M. Evangelopoulos, S. Acciardo, C. Corbo, E. De Rosa, I. Yazdi, S. Scaria, R. Molinaro, N. T. Furman *et al.*, "Biomimetic carriers mimicking leukocyte plasma membrane to increase tumor vasculature permeability," *Scientific reports*, vol. 6, 2016.

[48] Y. Qiu, M. Sridhar, J. K. Tsou, K. K. Lindfors, and M. F. Insana, "Ultrasonic viscoelasticity imaging of nonpalpable breast tumors: preliminary results," *Academic radiology*, vol. 15, no. 12, pp. 1526–1533, 2008.

[49] R. Righetti, J. Ophir, B. S. Garra, R. M. Chandrasekhar, and T. A. Krouskop, "A new method for generating poroelastograms in noisy environments," *Ultrasonic imaging*, vol. 27, no. 4, pp. 201–220, 2005.

[50] "regionprops," https://www.mathworks.com/help/images/ref/regionprops.html, 2017, accessed: 2017-10-14.

[51] A. P. Sarvazyan, O. V. Rudenko, S. D. Swanson, J. B. Fowlkes, and S. Y. Emelianov, "Shear wave elasticity imaging: a new ultrasonic technology of medical diagnostics," *Ultrasound in medicine & biology*, vol. 24, no. 9, pp. 1419–1435, 1998.

[52] J. Bercoff, M. Tanter, and M. Fink, "Supersonic shear imaging: a new technique for soft tissue elasticity mapping," *IEEE transactions on ultrasonics, ferroelectrics, and frequency control*, vol. 51, no. 4, pp. 396–409, 2004.

[53] E. Elyas, E. Papaevangelou, E. J. Alles, J. T. Erler, T. R. Cox, S. P. Robinson, and J. C. Bamber, "Correlation of ultrasound shear wave elastography with pathological analysis in a xenografic tumour model," *Scientific Reports*, vol. 7, 2017.

[54] A. Tahmasbi, F. Saki, and S. B. Shokouhi, "Classification of benign and malignant masses based on zernike moments," *Computers in biology and medicine*, vol. 41, no. 8, pp. 726–735, 2011.

[55] A. Boccaccini and G. Ondracek, "Dependence of the poisson number on the porosity in ceramic materials," *Cerám. Cristal*, no. 109, pp. 32–35, 1992.

[56] Y. Hassid, E. Furman-Haran, R. Margalit, R. Eilam, and H. Degani, "Noninvasive magnetic resonance imaging of transport and interstitial fluid pressure in ectopic human lung tumors," *Cancer research*, vol. 66, no. 8, pp. 4159–4166, 2006.

[57] R. K. Jain, "Transport of molecules in the tumor interstitium: a review," *Cancer research*, vol. 47, no. 12, pp. 3039–3051, 1987.

[58] M. S. Richards, P. E. Barbone, and A. A. Oberai, "Quantitative three-dimensional elasticity imaging from quasi-static deformation: a phantom study," *Physics in Medicine & Biology*, vol. 54, no. 3, p. 757, 2009.

[59] M. Tyagi, Y. Wang, T. Hall, P. E. Barbone, and A. A. Oberai, "Improving three-dimensional mechanical imaging of breast lesions with principal component analysis," *Medical physics*, 2017.

[60] M. Lekka, "Discrimination between normal and cancerous cells using afm," *Bionanoscience*, vol. 6, no. 1, pp. 65–80, 2016.

# Non-invasive imaging of Young's modulus and Poisson's ratio in cancers in vivo Supplementary information

**Islam et al.**

## 1. Validation of the proposed approach by finite element analysis

**Model.** The poroelastic sample containing a poroelastic inclusion used for the analysis reported in this paper is shown in Fig. S1 (A). In this figure, we see that the sample is of cylindrical shape. The inclusion can be of different shape (see Table I) as considered in the paper. However, we assumed that the shape of the inclusion is always axisymmetric, i.e., the shape remains the same if a 2D plane containing the inclusion is revolved around the center line. As an example, we show a spherical inclusion of radius $a$ inside the sample in Fig. S1 (A), which is perfectly axisymmetric. Because of the cylindrical symmetry of the sample and axisymmetry of the inclusion, the solution plane for this problem can be assumed as 2D as shown in Fig. S1 (B). From this figure, we also see that the compression is applied from the top and the bottom side is fixed. Two frictionless compressor plates have been used for holding up the sample and exert compression upon it.

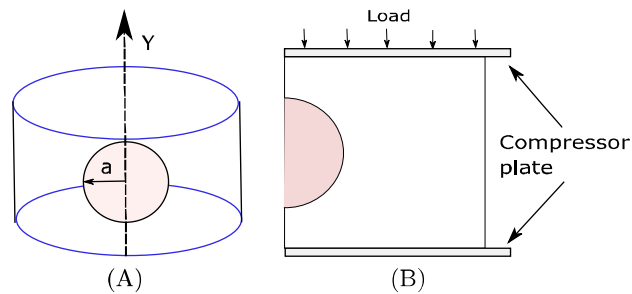

**Fig. S1.** (A) A schematic of a cylindrical sample of a poroelastic material with a spherical poroelastic inclusion of radius $a$. The axial direction is along the $y$-axis. (B) The 2D solution plane for the three dimensional sample. The sample is compressed between two compressor plates. The compression is applied along the negative $y$ direction.

**Finite element simulations.** A commercial finite element simulation software namely Abaqus, Dassault Systemes Simulia Corp., Providence, RI, USA has been used for the finite element simulation study. Both the inclusion and background of the sample are modeled as a linearly elastic, isotropic, incompressible, permeable solid phase saturated with an incompressible fluid.

The sample is compressed from the top and the bottom side is kept static. An instantaneous load of 1000 Pa is applied and kept constant for a certain time interval. This value of compression is chosen based on elastography experiments. The mesh used to model the sample is CAX4RP and has 81,790 elements in the solution plane. A zero fluid pressure boundary condition on the right hand side of the samples is imposed to ensure the flow of the fluid only in the right direction. We refer to Islam et al. (1) for details of the poroelastic simulation in Abaqus. The dimension of the solution plane of the sample is 2 cm in radius and 4 cm in height.

In all simulated samples, the interstitial permeability of the inclusion is taken as $3.1 \times 10^{-14}$ m$^2$ (Pa s)$^{-1}$ and of the background is taken as $6.4 \times 10^{-15}$ m$^2$ (Pa s)$^{-1}$. The vascular permeability of the inclusion is assumed $5.67 \times 10^{-7}$ (Pa s)$^{-1}$ and of the background is assumed $1.89 \times 10^{-8}$ (Pa s)$^{-1}$. These values of material properties for the inclusion and background are chosen following Leiderman et al. (2). Void ratio in all the samples is assumed equal to 0.4. These material properties are kept the same in all the samples (A-M, Z1-Z8, X1-X9, H1-H3, B1-B3 and R1-R4) while the values of YM and PR are varied to create different simulation conditions. In the samples that have non-uniform background, i.e., small inclusions, stiff strip of tissue, etc. the values of interstitial and vascular permeabilities for these objects in the background are assumed the same as those for the background. The analysis time in Abaqus is chosen for different samples in such a way that the samples reach steady state and behave as fully linear elastic materials at the end of the analysis. The axial and lateral strains used in all the YM and PR reconstructions reported in the paper are computed at the steady state.

For segmenting the steady state axial and lateral stain elastograms from finite element analysis (FEA), a morphological segmentation algorithm is used (3).

**Reconstruction of YM and PR from FEA data of samples A-M.** The radius of the spherical inclusion in samples A-M is 0.3 cm. The mechanical parameters of samples A-M are tabulated in Table S1. In this table, $E_b$ and $E_i$ denote the YM of the normal tissue (background) and tumor (inclusion) and $\nu_b$ and $\nu_i$ denote their PR.

**Table S1. YM and PR of samples A-M used in the FEA and ultrasound simulations**

| Sample name | $E_b$ (kPa) | $E_i$ (kPa) | $\nu_b$ | $\nu_i$ |
|---|---|---|---|---|
| A | 32.78 | 97.02 | 0.49 | 0.40 |
| B | 32.78 | 50.00 | 0.49 | 0.40 |
| C | 32.78 | 163.90 | 0.49 | 0.40 |
| D | 32.78 | 97.02 | 0.45 | 0.45 |
| E | 32.78 | 97.02 | 0.20 | 0.45 |
| F | 32.78 | 97.02 | 0.20 | 0.30 |
| G | 32.78 | 491.70 | 0.20 | 0.45 |
| H | 32.08 | 819.50 | 0.20 | 0.45 |
| I | 32.78 | 1639.00 | 0.20 | 0.45 |
| J | 32.78 | 3278.00 | 0.20 | 0.45 |
| K | 32.08 | 16.39 | 0.20 | 0.30 |
| L | 32.78 | 6.556 | 0.20 | 0.30 |
| M | 32.78 | 3.278 | 0.20 | 0.30 |

The steady state axial and lateral strains for the first four samples (A-D) from FEA are shown in Figs. S2 (A1-A4) and (B1-B4), respectively. In this figures, we see that the axial and lateral strains are constant inside the inclusion, which correlates with Eshelby's theory (4).

The reconstructed YM of samples A-D by the proposed method and the other two methods used for the comparison are shown in Fig. S3 (A1-A4), (B1-B4) and (C1-C4), respectively. The reconstructed PR of samples A-D by the proposed method are shown in Fig. S4 (A1-A4).

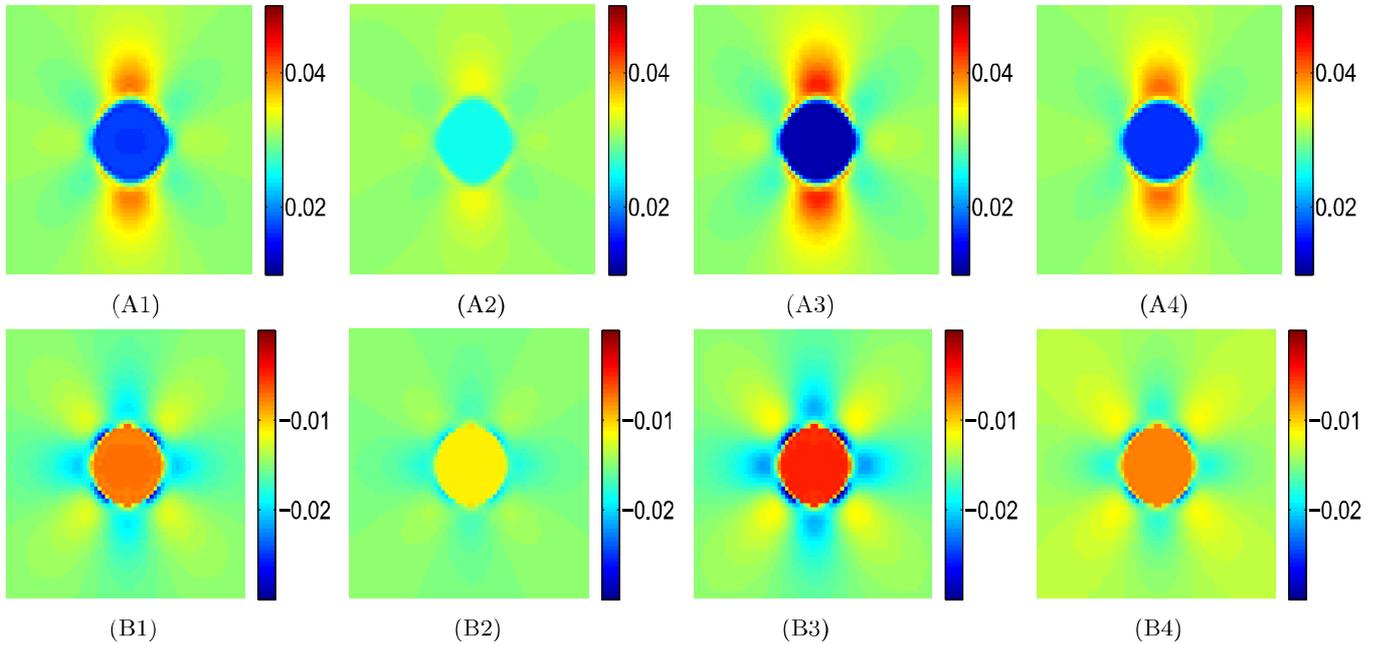

(A1) (A2) (A3) (A4)

(B1) (B2) (B3) (B4)

**Fig. S2.** (A1)-(A4) Axial strains from finite element simulation for samples A-D (B1)-(B4) Lateral strains from finite element simulation for samples A-D.

**Specifications of the samples Z1-Z8 and X1-X9.** For samples Z1-Z8, the YM and PR of tumor are set to 97.02 kPa and 0.3, whereas the YM and PR of the normal tissue are set to 32.78 kPa and 0.2. For samples X1-X9, the YM and PR of tumor are set to 97.02 kPa and 0.45 and the YM and PR of the normal tissue are set to 32.78 kPa and 0.2.

All the samples simulated are of 4 cm height and 2 cm width in an axisymmetric setup. In samples X1-X9, the radius of the spherical inclusion is 0.3 cm. In sample Z1, the radius of the inclusion is 0.3 cm, the lengths of elliptical axes along lateral and axial direction in inclusions of samples Z2 and Z3 are 0.2 cm and 0.5 cm and 0.5 cm and 0.2 cm, respectively. The radius and height of the cylindrical inclusion of sample Z4 are 0.3 cm and 0.55 cm. The radius of the penny-shaped inclusion of sample Z5 is 0.5 cm and the height is 0.05 cm. The length of each side of tetragonal, pentagonal and hexagonal inclusions in samples Z6, Z7 and Z8 are 0.45 cm.

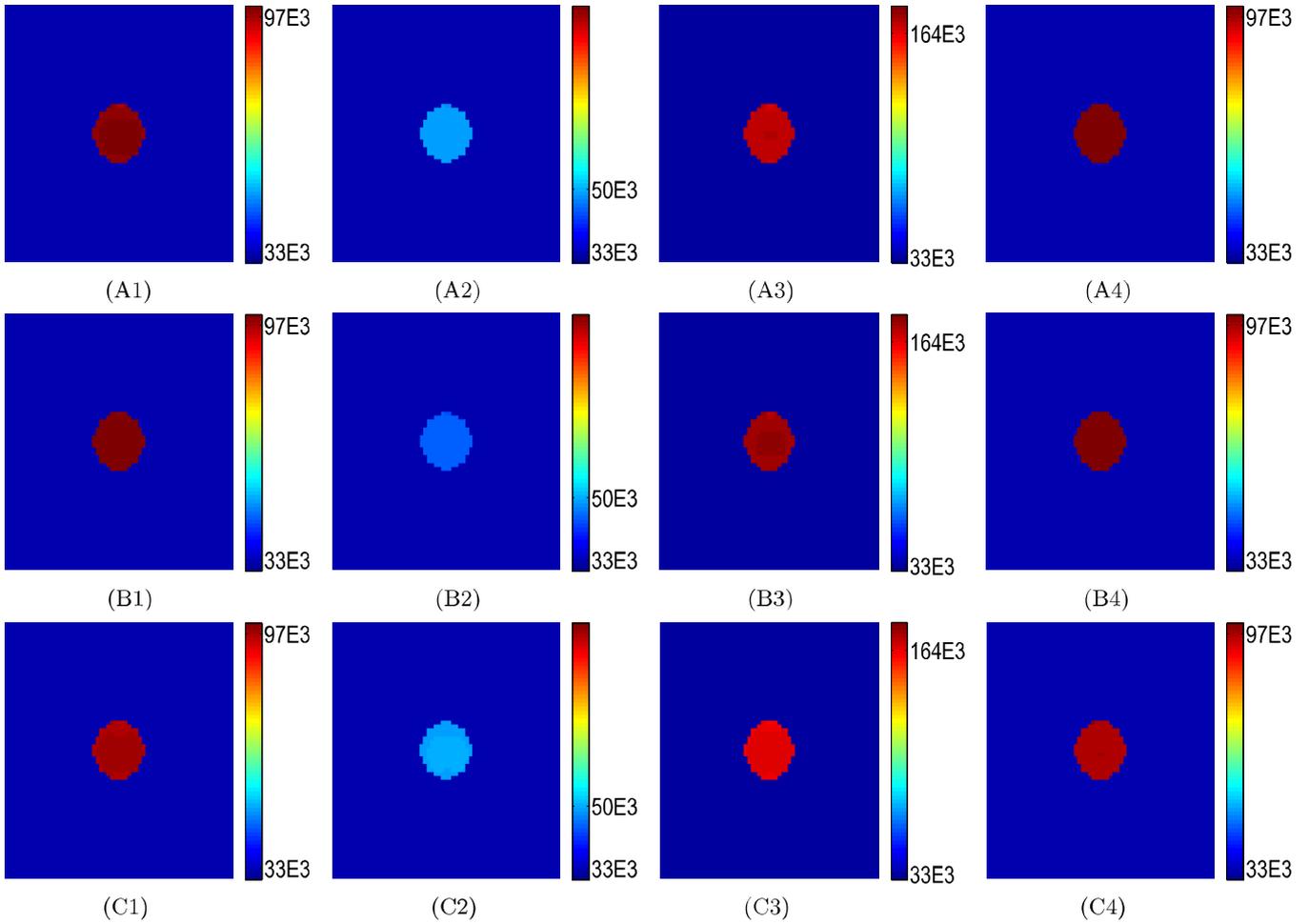

**Fig. S3.** (A1)-(A4) Reconstructed YM (in Pa) of samples A-D by the 3DB approach and (B1)-(B4) reconstructed YM (in Pa) of samples A-D by the 3DS approach from FEA axial strain data. (C1)-(C4) Reconstructed YM (in Pa) of samples A-D by the proposed approach from FEA axial and lateral strain data.

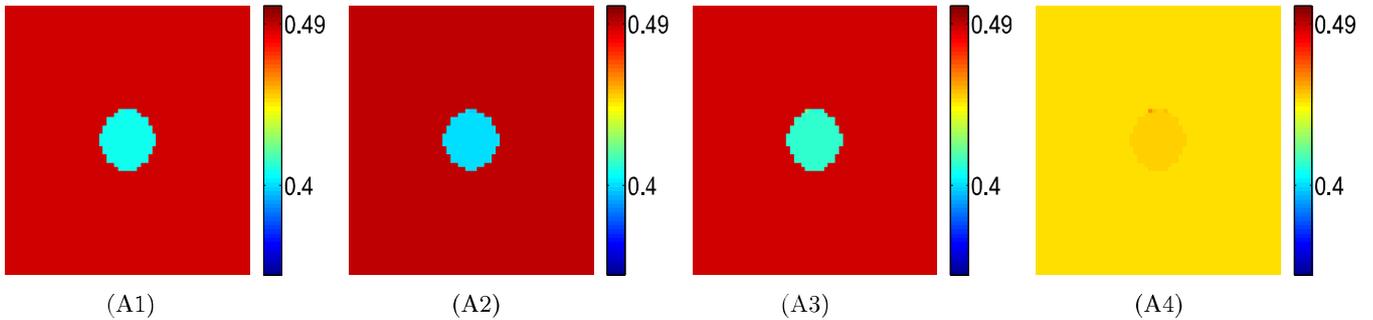

**Fig. S4.** (A1)-(A4) Reconstructed PR of samples A-D by the proposed approach from FEA axial and lateral strain data.

## 2. Finite element models of samples with complex boundary conditions

**A. Samples with zig-zag stiff tissue, spheres and ellipses in the background.** We have chosen two samples B1 and B2 with two different complex boundary conditions. In sample B1, we simulate zig-zag stiff materials (YM of 40 kPa) in the normal tissue region. In sample B2, there are fourteen spherical and elliptical inclusions in the normal tissue region with different YM (45 kPa, 50 kPa and 60 kPa) and PR (0.15, 0.2, 0.22, 0.25 and 0.3). The YM in the inclusion and normal tissue is 97.02 and 32.78 kPa, respectively in sample B1 and B2. In both samples, the main inclusion has a radius of 1 cm. The lowest distance from the stress application plane to the inclusion is 1 cm. As shown in the FEA model (Fig. S5), the total area of the sample considered is 8 cm× 4 cm. This area is chosen based on the size of commonly used gelpad (9 cm diameter) and compressor plate (10.1 cm × 9.3 cm) and the depth of ultrasound penetration used in common experiments which is 4 cm. The imaging region is the center square portion of 4 cm× 4 cm.

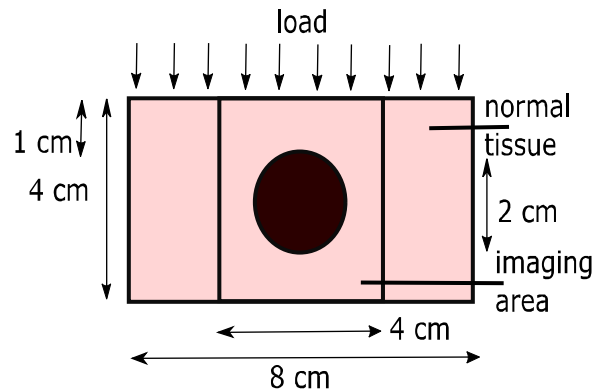

**Fig. S5.** Finite element model to investigate the impact of complex boundary conditions on the estimation of YM and PR by the proposed method.

**B. Sample with a strip of stiff tissue above the tumor.** The finite element model of the sample B3 with a strip of stiff tissue above the tumor is shown in Fig. S6. The YM in the tumor is 97.02 kPa and in the normal tissue is 32.78 kPa. YM in the strip stiff tissue is 50 kPa. PR in the tumor is assumed as 0.3 and in the normal tissue and stiff tissue strip as 0.2. The radius of the spherical inclusion inside sample B3 is 0.3 cm. The height and width of sample B3 are 4 cm and 2 cm, respectively.

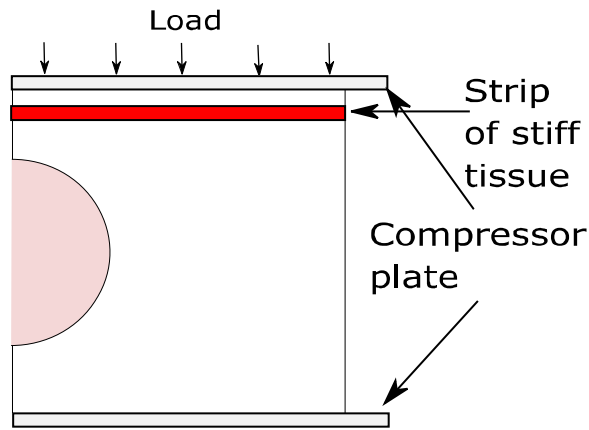

**Fig. S6.** Finite element model of samples B3 with a strip of stiff tissue above the spherical tumor.

## 3. Samples with a heterogeneous YM distribution inside the tumor

The finite element models for samples H1, H2 and H3 with heterogeneous YM distributions inside the tumor are shown in Fig. S7. The inclusion in samples H1-H3 is made up of three concentric spheres of outer radii 0.1, 0.2 and 0.3 cm. The inner sphere has YM modulus of 97.02 kPa in samples H1-H3 and the next spherical shell has YM of 92.17 kPa for 10% heterogeneity (sample H1), 87.32 kPa for 20% (sample H2) and 82.46 kPa for 30% heterogeneity (sample H3). The peripheral spherical shell has YM of 87.32 kPa for 10% heterogeneity, 77.62 kPa for 20% and 67.91 kPa for 30% heterogeneity. The YM of the normal tissue is assumed to be 32.78 kPa. The PR of the tumor is taken as 0.3 and of the normal tissue is taken as 0.2. The height and width of samples H1-H3 are 4 cm and 2 cm, respectively.

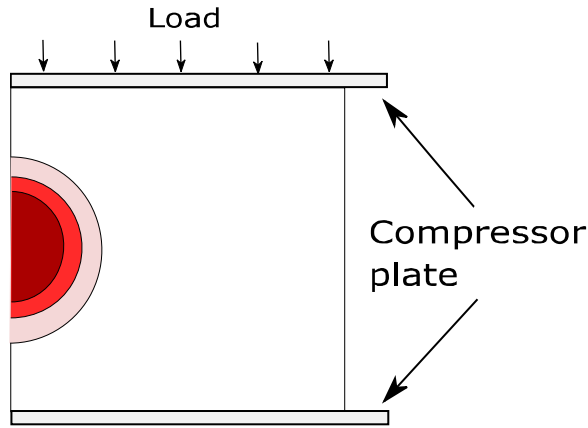

**Fig. S7.** Finite element model of samples H1, H2 and H3 with heterogeneous YM distribution inside the tumor.

## 4. Finite element model of samples subjected to non-uniform compression

Among the simulated samples subjected to non-uniform compressions (shown in Fig. S8) in FEA, in sample R1, the load is 1 kPa in the center, which gets reduced by 20% at the side of the imaging area. In sample R2, the load is 1 kPa in the center, which gets reduced by 10% at the side of the imaging area. In sample R3, the load is 1 kPa in the center which increases by 20% at the side of the imaging area. In sample R4, the load is 1 kPa in the center which increases by 10% at the side of the imaging area. In samples R1-R4, the YM and PR of the inclusion are set to 97.02 kPa and 0.3, whereas the YM and PR of the background are set to 32.78 kPa and 0.2.

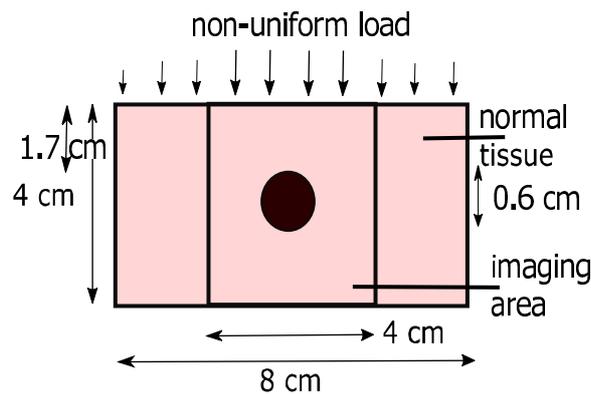

**Fig. S8.** Finite element model to investigate the impact of non-uniform compression on the estimation of YM and PR by the proposed method.

## 5. Finite element model of samples with multiple layers above the inclusion

The FEA model for samples with multiple layers of soft tissue with different stiffness is shown in Fig. S9. We chose the YM of the four layers as 30 kPa, 49.17 kPa, 65.56 kPa and 81.95 kPa in sample B4 and B5. We selected 32.78 kPa as the YM of the background tissue. In both the layer tissues and normal tissues, we assumed PR of 0.4 in sample B4, whereas we assumed PR of 0.25 in sample B5. The YM of the tumor has been assumed 97.02 and 163.90 kPa in sample B4 and B5, respectively. The PR of the tumor has been assumed as 0.4 in sample B4 and 0.3 in sample B5.

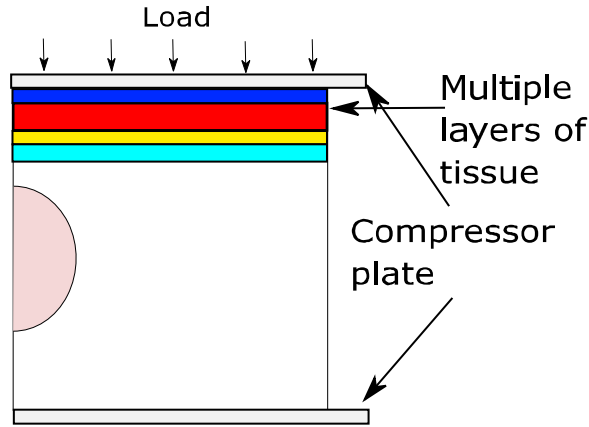

**Fig. S9.** Finite element model of samples B4-B5 with four layers of tissue of different YM above the spherical tumor.

## 6. Ultrasound simulation

**Simulation method.** The simulated pre- and post-compression ultrasound RF data are generated from the mechanical displacements using a convolution model (5). Bilinear interpolation is performed on the input mechanical displacement data (obtained from FEA) prior to the computation of the simulated RF frames (6). The simulated ultrasound transducer has 128 elements, frequency bandwidth between $5-14$ MHz, a 6.6 MHz center frequency, and 50% fractional bandwidth at $-6$ dB. The transducer's beamwidth is assumed to be dependent on the wavelength and to be approximately 1 mm at 6.6 MHz (7). The sampling frequency is set at 40 MHz and Gaussian noise is added to set the SNR at 40 dB. From the same sets of simulated pre- and post-compression RF data, the method proposed by Islam et al. (8) is used to estimate the axial and lateral strains. Axial and lateral strains from 50 independent realizations are averaged to obtain the final axial and lateral strains.

For segmenting the axial and lateral stain elastograms from ultrasound simulated data, a morphological segmentation algorithm is used as in (3).

**Comparison.** The axial and lateral strains for ultrasound simulations for the first four samples (A-D) are shown in (A1)-(A4) and (B1-B4) of Fig. S10.

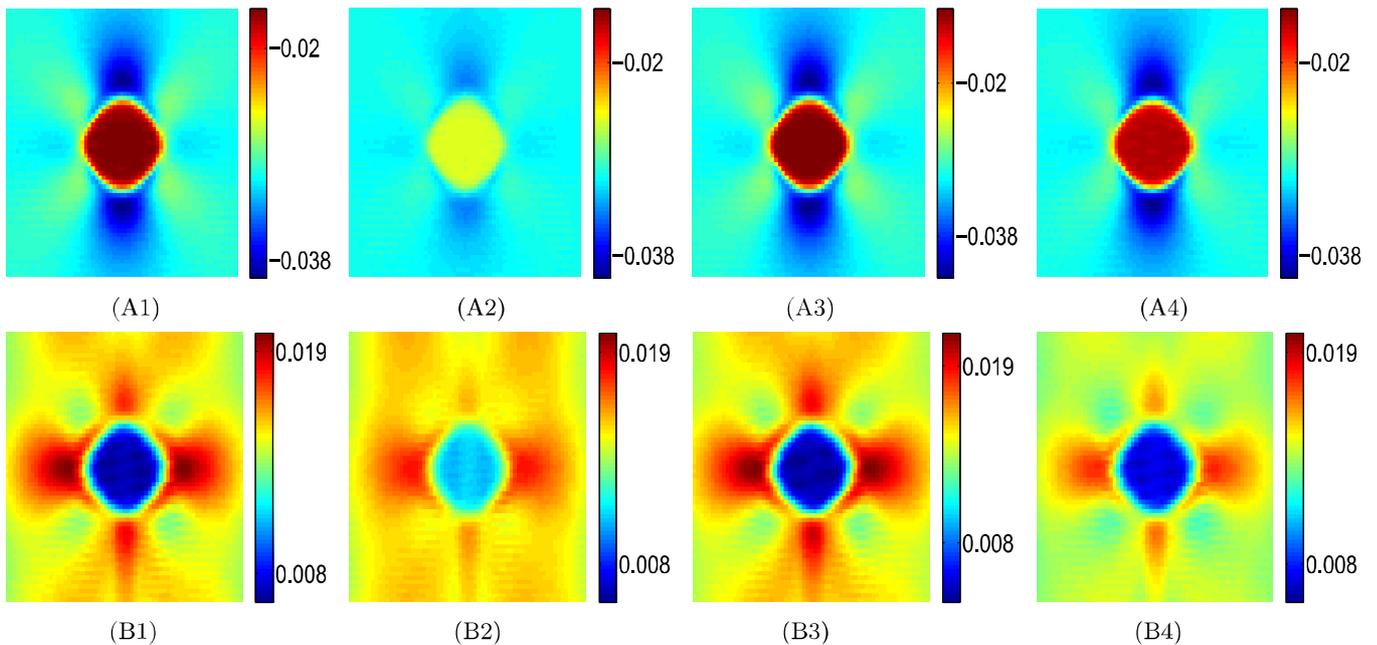

**Fig. S10.** (A1)-(A4) Estimated axial strains from ultrasound simulation data for samples A-D (B1)-(B4) estimated lateral strains from ultrasound simulation data for samples A-D.

The estimated values of YM by the 3DB approach are shown in (A1-A4) of Fig. S11 for ultrasound simulated RF data of samples A-D. The estimated values of YM appear uniform inside the inclusion, but the values significantly deviate from the true values. The reconstructed YM by the 3DS approach are shown in (B1-B4) of Fig. S11 for ultrasound simulated RF data of samples A-D. The reconstructed YM and PR by our proposed approach are shown in (C1-C4) of Fig. S11 and (A1-A4) of Fig. S12. The estimated YM and PR by our technique are closer to the true values than the estimated YM and PR estimated by other two methods.

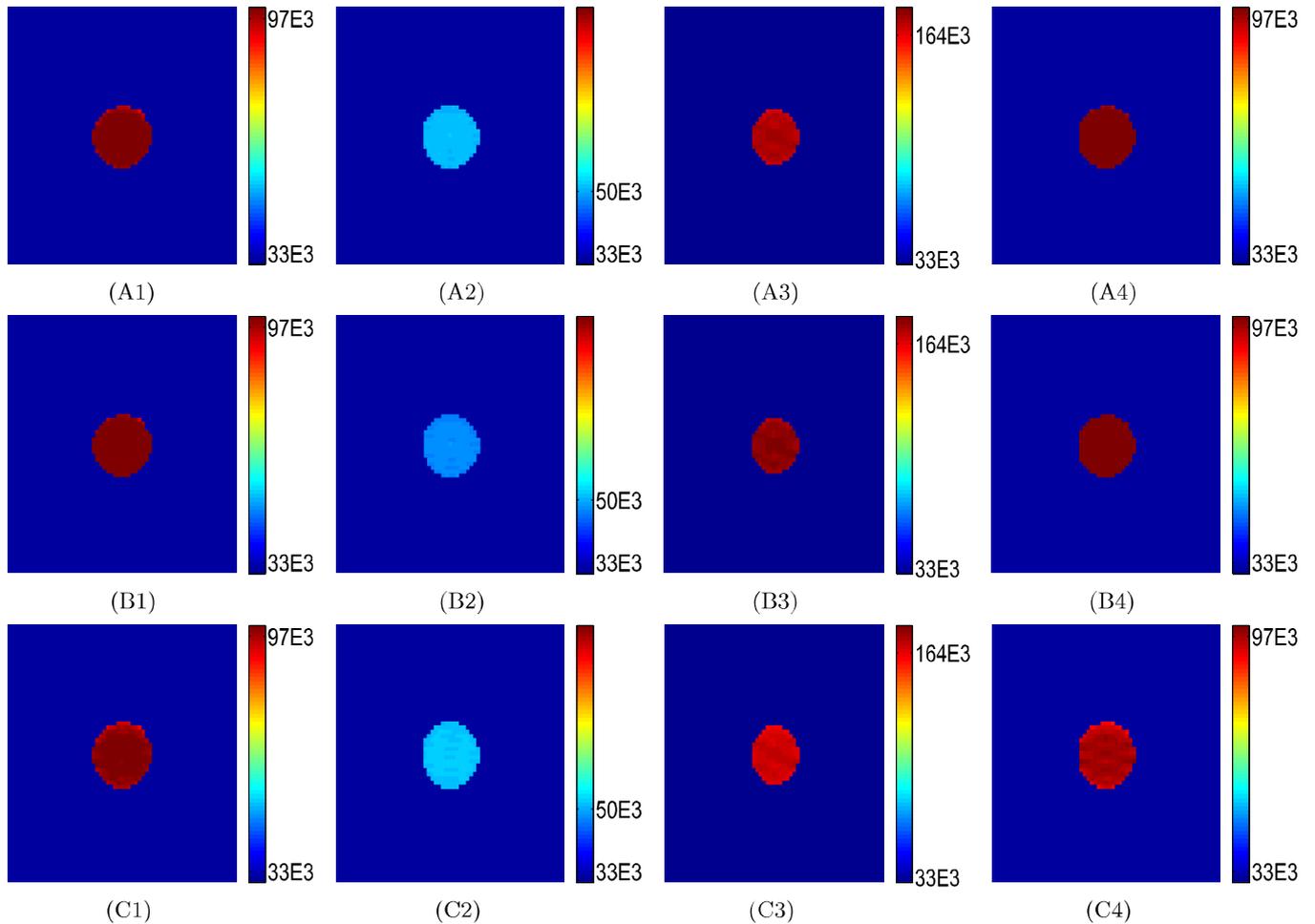

**Fig. S11.** (A1)-(A4) Reconstructed YM (in Pa) in samples A-D by the 3DB approach as applied to ultrasound simulated RF data. (B1)-(B4) Reconstructed YM (in Pa) in samples A-D by the 3DS approach and (C1)-(C4) reconstructed YM (in Pa) in samples A-D by the proposed method as applied to ultrasound simulated RF data.

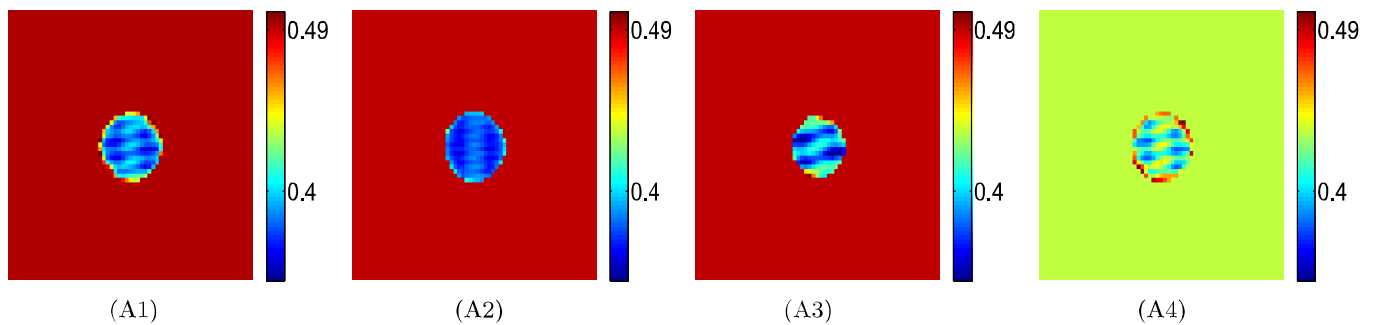

**Fig. S12.** (A1)-(A4) Reconstructed PR in samples A-D using the proposed approach as applied to ultrasound simulated RF data.

## 7. Axial strain ratios for different YM contrasts

The ratio of the axial strains inside and outside the inclusion estimated using Eshelby's theory (4) and its logarithm are shown in Fig. S13 (A) and (B), respectively for a range of YM ratio (0.0001 − 10000) of the inclusion and background. We observe from Fig. S13 (A) that the strain ratio decreases as the YM ratio between the inclusion and background increases above 1, and the strain ratio increases as the YM ratio between the inclusion and background decreases below 1. Below and above certain YM ratios, the strain ratio is saturated. Although the strain curve seems symmetric with respect to value of 1, the ratio of two strain ratio values is much smaller in the 1 − 2 region than in the 0 − 1 region. This becomes clearer if the logarithm of the strain ratio is taken, as shown in Fig. S13 (B). For the hard tumors (the region in the right side of 0 along x axis), we see that the logarithm of the strain ratio changes along a sloped straight line with respect to the modulus ratio. Therefore, two values of strain ratio in this region are distinct. For the soft tissue tumors (the region in the left side of 0 along x axis), the logarithm of strain ratio becomes almost constant when the modulus ratio is less than 0.1. Below the modulus ratio of 0.1, it becomes harder for an algorithm to differentiate between two distinct strain ratio values and as a result to determine the actual YM of the tumor. This is the reason why the determination of YM of soft tumors is more challenging in comparison to the hard ones. Although we see that the logarithm of the strain ratio is changing in the hard tumor region even for modulus ratio values above 100, in practice, the difference between two strain ratio values in this region in linear scale is very small to be detected by the reconstruction algorithms, especially when noisy ultrasound strain data are used for the reconstruction.

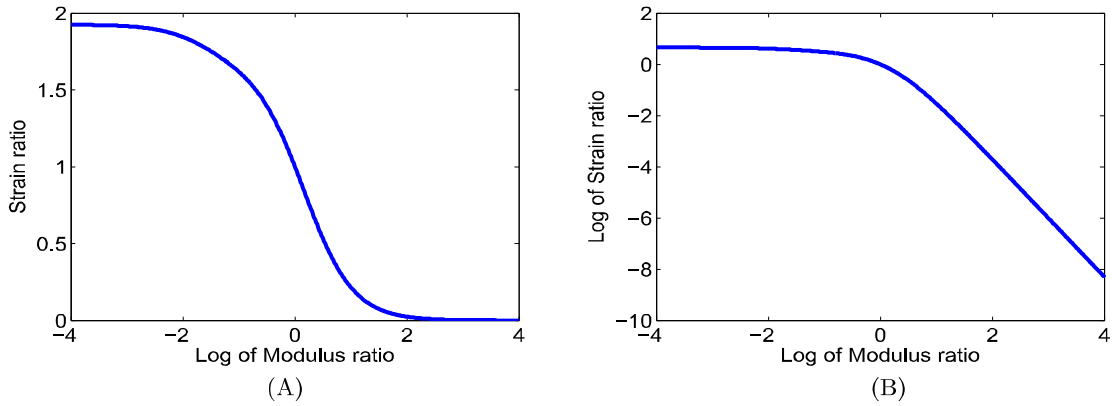

**Fig. S13.** (A) Strain ratio (ratio of axial strains of the inclusion and background) and (B) log of strain ratio for different Young's modulus ratio.

## 8. Illustration of stress decay in the our experimental setup

When a compressor of finite size is used to excite a tissue, the applied stress is the highest near the top boundary and decreases with depth in the sample (9, 10). This 'stress decay' leads to an overestimation of the reconstructed YM in the bottom part of the sample. This problem has been previously observed in elastography and has been referred to as 'target hardening effect' (11, 12). Using our current experimental setup, we found this effect to be insignificant on the reconstructed results as proven below.

A finite element model to investigate the impact of stress decay on the observed strains in an elastography experiment due to the finite size of the compressor is shown in Fig. S14 (A). In this simulation, the YM and PR of the uniform sample are taken as 32.78 kPa and 0.2, respectively. A uniaxial load of 1 kPa has been applied from the top. The decayed axial stress in the imaging area is shown in Fig. S14 (B). We see from Fig. S14 (B) that even in the bottom part of the imaging area, the stress reduction is not larger than 2%. As, in this simulation, the stress distribution inside the sample when a finite sized compressor is used is similar to the stress distribution inside the sample when an infinite sized compressor is used (stress is equal to the applied stress inside the entire sample), the observed strains in our elastography experiments would be almost identical to the theoretical strains computed assuming an infinite sized compressor.

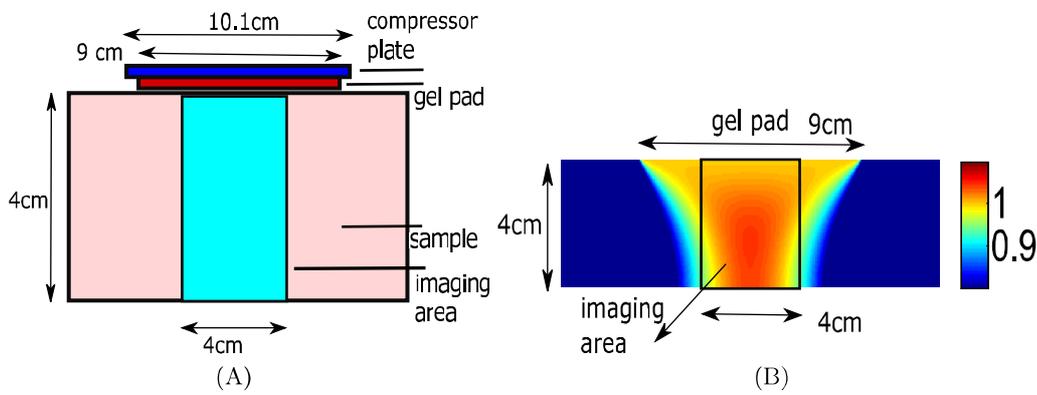

**Fig. S14.** (A) Finite element model to investigate the impact of stress decay on the observed strains in our experimental setup (B) axial stress (in kPa) in the imaging area of our setup.

## 9. Stress distribution in samples with gel pad

The finite element model for simulating the stress and strain distributions in samples with a gel pad at the top is shown in Fig. S15. The YM and PR of normal tissue have been assumed 32.78 kPa and 0.4. The YM and PR of the tumor have been assumed 97.02 kPa and 0.4. The YM of the soft gelpad has been assumed 16.39 kPa and of the stiff gelpad has been assumed 65.56 kPa. The PR of both type of gelpad has been assumed 0.4. A load of 1 kPa has been applied on the top surface in this model. The stress distributions for the stiff and soft gel pads are shown in Fig. S16. We show the mean axial and lateral stresses in all three cases in Table S2. We see from this table that the axial and lateral stresses change insignificantly because of the gelpad usage. Based on these simulation results, it can be said that the observed strains in the tissues in our elastography experiments are not affected by the presence of the gelpad.

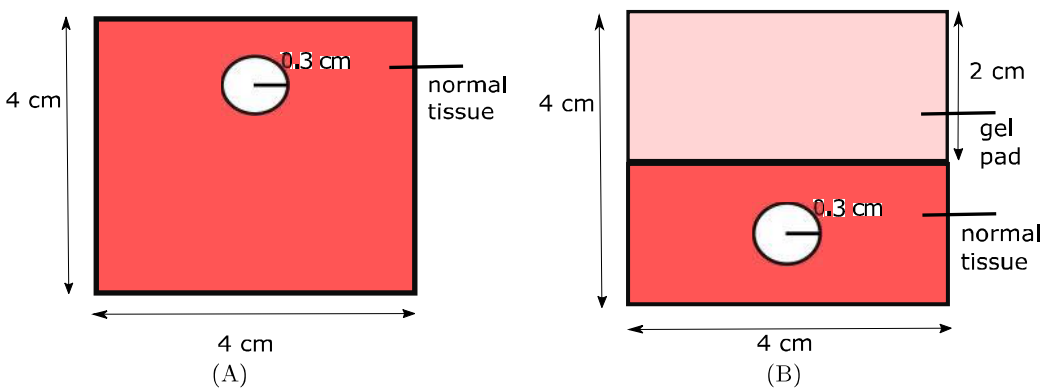

**Fig. S15.** Finite element model of a non-uniform sample with gelpad at the top used to investigate the impact of the use of gelpad on the observed strains in our in vivo tissues. Sample (A) without gelpad (B) with gelpad.

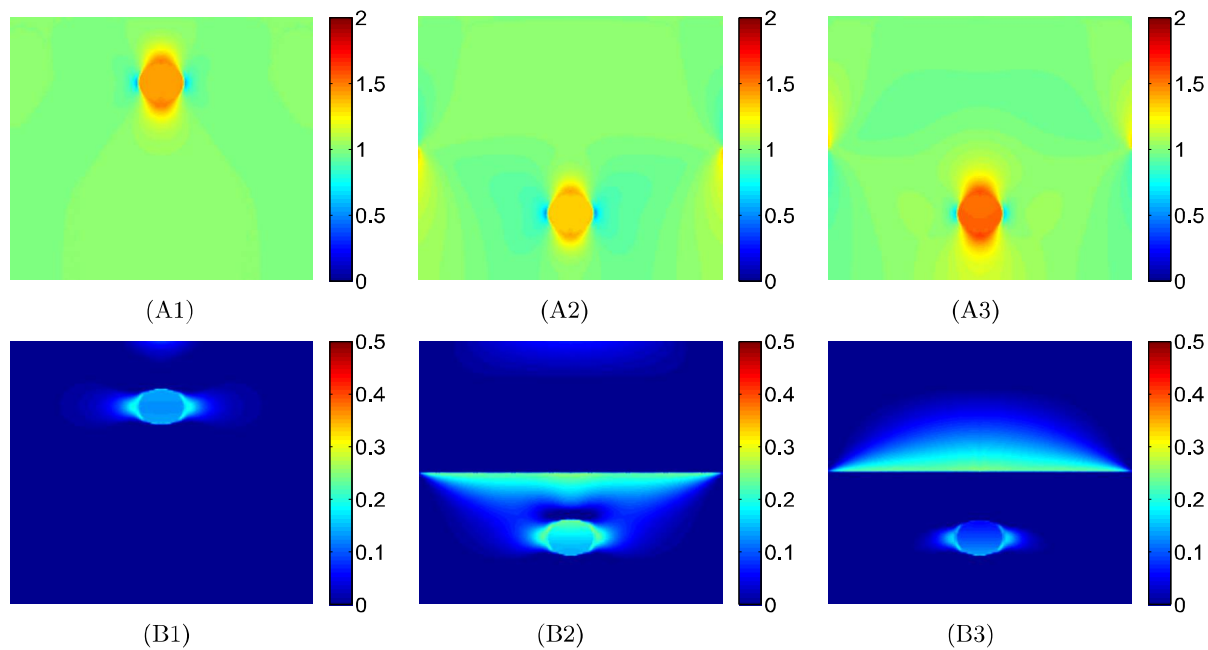

**Fig. S16.** Axial stress (in kPa) (A1) without gelpad (A2) with 50% softer and (A3) 200% stiffer gelpad with respect to normal tissue. Lateral stress (in kPa) (B1) without gelpad (B2) with 50% softer and (B3) 200% stiffer gelpad with respect to normal tissue.

**Table S2. Mean axial and lateral stresses inside the inclusions for samples with and without gelpad**

|  | Axial stress (kPa) | Lateral stress (kPa) |
|---|---|---|
| Sample without gelpad | 1.43 | 0.13 |
| Sample with soft gelpad | 1.33 | 0.16 |
| Sample with stiff gelpad | 1.53 | 0.09 |

## 10. Eshelby's virtual experiment

The virtual experiment is composed by the following steps:

1. Isolate the inclusion from the background (Fig. S17(A)). Consequently, the inclusion is strained because of the loss of constraint imposed by the background. This strain is denoted as eigenstrain ($\epsilon^*$).

2. Apply traction $T$ to bring the inclusion in its original shape (Fig. S17(B)). The strain induced inside the inclusion should compensate the eigenstrain.

3. Insert the inclusion back in the background (Fig. S17(C)). The traction force is still $T$.

4. Remove the applied traction $T$ (Fig. S17(D)). This is the same scenario as step 1 (Fig. S17(A)). The removal of the traction force from step 3 to step 4 is equivalent to applying a body force of $-T$ to the surface of the inclusion.

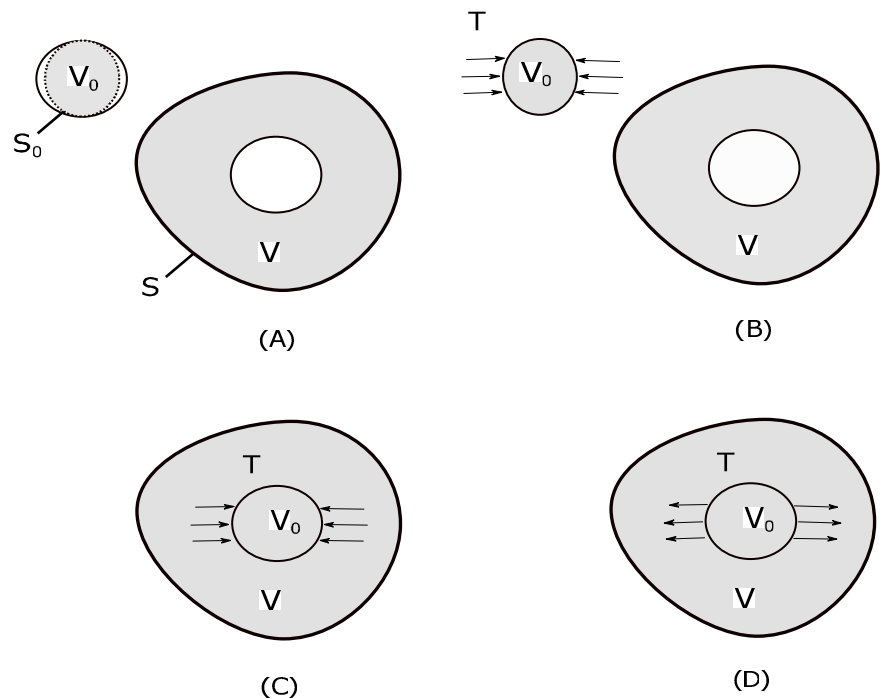

**Fig. S17.** Four steps of Eshelby's virtual experiment to reach the solution. Here, background is a linear elastic solid of volume $V$ and surface $S$. The inclusion is also a linear elastic solid of volume $V_0$ and surface $S_0$. (A) The inclusion is removed from the background. (B) A surface traction $T$ is applied to return $V_0$ in its original shape. (C) We put the inclusion back to the matrix and (D) remove the applied traction.

Islam et al. 11 of 26

## 11. Expression of $A$

The stiffness matrix for the inclusion can be written as

$$C = \begin{bmatrix} \lambda_i + 2\mu_i & \lambda_i & \lambda_i & 0 & 0 & 0 \\ \lambda_i & \lambda_i + 2\mu_i & \lambda_i & 0 & 0 & 0 \\ \lambda_i & \lambda_i & \lambda_i + 2\mu_i & 0 & 0 & 0 \\ 0 & 0 & 0 & \mu_i & 0 & 0 \\ 0 & 0 & 0 & 0 & \mu_i & 0 \\ 0 & 0 & 0 & 0 & 0 & \mu_i \end{bmatrix}, \qquad [1]$$

where

$$\lambda_i = \frac{E_i \nu_i}{(1+\nu_i)(1-2\nu_i)}, \qquad [2]$$

$$\mu_i = \frac{E_i}{2(1+\nu_i)}. \qquad [3]$$

Here $E_i$ and $\nu_i$ are the YM and PR of the inclusion. The stiffness matrix for the background can be written as

$$C^0 = \begin{bmatrix} \lambda_b + 2\mu_b & \lambda_b & \lambda_b & 0 & 0 & 0 \\ \lambda_b & \lambda_b + 2\mu_b & \lambda_b & 0 & 0 & 0 \\ \lambda_b & \lambda_b & \lambda_b + 2\mu_b & 0 & 0 & 0 \\ 0 & 0 & 0 & \mu_b & 0 & 0 \\ 0 & 0 & 0 & 0 & \mu_b & 0 \\ 0 & 0 & 0 & 0 & 0 & \mu_b \end{bmatrix}, \qquad [4]$$

where

$$\lambda_b = \frac{E_b \nu_b}{(1+\nu_b)(1-2\nu_b)}, \qquad [5]$$

$$\mu_b = \frac{E_b}{2(1+\nu_b)}. \qquad [6]$$

Here $E_b$ and $\nu_b$ are the YM and PR of the background.

If we assume that $X = (C - C^0)^{-1}$, then $X$ can be written as

$$X = \begin{bmatrix} \Psi & \Omega & \Omega & 0 & 0 & 0 \\ \Omega & \Psi & \Omega & 0 & 0 & 0 \\ \Omega & \Omega & \Psi & 0 & 0 & 0 \\ 0 & 0 & 0 & -1/(\mu_b - \mu_i) & 0 & 0 \\ 0 & 0 & 0 & 0 & -1/(\mu_b - \mu_i) & 0 \\ 0 & 0 & 0 & 0 & 0 & -1/(\mu_b - \mu_i) \end{bmatrix}, \qquad [7]$$

where

$$\Psi = -(\lambda_b - \lambda_i + \mu_b - \mu_i)/(2\mu_b^2 + 2\mu_i^2 + 3\lambda_b\mu_b - 3\lambda_b\mu_i - 3\lambda_i\mu_b + 3\lambda_i\mu_i - 4\mu_b\mu_i), \qquad [8]$$

$$\Omega = (\lambda_b - \lambda_i)/(2(2\mu_b^2 + 2\mu_i^2 + 3\lambda_b\mu_b - 3\lambda_b\mu_i - 3\lambda_i\mu_b + 3\lambda_i\mu_i - 4\mu_b\mu_i)). \qquad [9]$$

The matrix $A = [C - C^0]^{-1} \cdot C^0$ can be written as

$$A = \begin{bmatrix} \Psi(\lambda_b + 2\mu_b) + 2\lambda_b\Omega & \Omega(\lambda_b + 2\mu_b) + \lambda_b\Omega + \lambda_b\Psi & \Omega(\lambda_b + 2\mu_b) + \lambda_b\Omega + \lambda_b\Psi & 0 & 0 & 0 \\ \Omega(\lambda_b + 2\mu_b) + \lambda_b\Omega + \lambda_b\Psi & \Psi(\lambda_b + 2\mu_b) + 2\lambda_b\Omega & \Omega(\lambda_b + 2\mu_b) + \lambda_b\Omega + \lambda_b\Psi & 0 & 0 & 0 \\ \Omega(\lambda_b + 2\mu_b) + \lambda_b\Omega + \lambda_b\Psi & \Omega(\lambda_b + 2\mu_b) + \lambda_b\Omega + \lambda_b\Psi & \Psi(\lambda_b + 2\mu_b) + 2\lambda_b\Omega & 0 & 0 & 0 \\ 0 & 0 & 0 & \mu_b\Phi & 0 & 0 \\ 0 & 0 & 0 & 0 & \mu_b\Phi & 0 \\ 0 & 0 & 0 & 0 & 0 & \mu_b\Phi \end{bmatrix}, \qquad [10]$$

where

$$\Phi = -\frac{1}{\mu_b - \mu_i}. \qquad [11]$$

## 12. Expressions of $\epsilon_1^*$ and $\epsilon_2^*$ for elliptic inclusion

The Eshelby's tensor $S$ can be written for an elliptic inclusion with semi-axis lengths of $a, b$ and $c$ along $x, y$ and $z$-direction as (13)

$$S = \begin{bmatrix} S_{1111} & S_{1122} & S_{1133} & 0 & 0 & 0 \\ S_{2211} & S_{2222} & S_{2233} & 0 & 0 & 0 \\ S_{3311} & S_{3322} & S_{3333} & 0 & 0 & 0 \\ 0 & 0 & 0 & S_{2323} & 0 & 0 \\ 0 & 0 & 0 & 0 & S_{3131} & 0 \\ 0 & 0 & 0 & 0 & 0 & S_{1212} \end{bmatrix}, \quad [12]$$

where

$$S_{1111} = \frac{3}{8\pi(1-\nu_b)} a^2 I_{11} + \frac{1-2\nu_b}{8\pi(1-\nu_b)} I_1, \quad [13]$$

$$S_{1122} = \frac{1}{8\pi(1-\nu_b)} b^2 I_{12} + \frac{1-2\nu_b}{8\pi(1-\nu_b)} I_1, \quad [14]$$

$$S_{1133} = \frac{1}{8\pi(1-\nu_b)} c^2 I_{13} + \frac{1-2\nu_b}{8\pi(1-\nu_b)} I_1, \quad [15]$$

$$S_{1212} = \frac{a^2+b^2}{16\pi(1-\nu_b)} a^2 I_{12} + \frac{1-2\nu_b}{16\pi(1-\nu_b)} (I_1 + I_2). \quad [16]$$

The other nonzero terms can be found by cyclic permutation of the above formulas. We have to let $a \to b \to c$ together with $1 \to 2 \to 3$.

Assuming $a > b > c$, the $I$-terms can be calculated as

$$I_1 = \frac{4\pi abc}{(a^2-b^2)(a^2-c^2)^{\frac{1}{2}}} \big[ \chi(\theta, k) - \zeta(\theta, k) \big], \quad [17]$$

$$I_3 = \frac{4\pi abc}{(b^2-c^2)(a^2-c^2)^{\frac{1}{2}}} \Big[ \frac{b(a^2-c^2)^{\frac{1}{2}}}{ac} - \zeta(\theta, k) \Big], \quad [18]$$

where

$$\theta = \sin^{-1} \sqrt{\frac{a^2-c^2}{a^2}}, \quad [19]$$

$$k = \sqrt{\frac{a^2-b^2}{a^2-c^2}} \quad [20]$$

and

$$I_1 + I_2 + I_3 = 4\pi, \quad [21]$$

$$3I_{11} + I_{12} + I_{13} = \frac{4\pi}{a^2}, \quad [22]$$

$$3a^2 I_{11} + b^2 I_{12} + c^2 I_{13} = 3I_1, \quad [23]$$

$$I_{12} = \frac{I_2 - I_1}{a^2 - b^2}. \quad [24]$$

The standard elliptic integrals $\chi$ and $\zeta$ are defined as

$$\chi(\theta, k) = \int_0^\theta \frac{dw}{\sqrt{1-k^2 sin^2 w}}, \quad [25]$$

$$\zeta(\theta, k) = \int_0^\theta \sqrt{1-k^2 sin^2 w}\, dw. \quad [26]$$

The eigen strain can be written as (4, 14)

$$\epsilon_1^* = S^{-1}[\epsilon - \epsilon^0], \quad [27]$$

where the background and inclusion strain can be expressed as

$$\boldsymbol{\epsilon^0} = \begin{bmatrix} \epsilon_{11}^0 \\ \epsilon_{22}^0 \\ \epsilon_{33}^0 \\ 0 \\ 0 \\ 0 \end{bmatrix} \qquad [28]$$

and

$$\boldsymbol{\epsilon} = \begin{bmatrix} \epsilon_{11} \\ \epsilon_{22} \\ \epsilon_{33} \\ 0 \\ 0 \\ 0 \end{bmatrix}. \qquad [29]$$

Another expression of the eigen strain can be written as (15, 16)

$$\boldsymbol{\epsilon_2^*} = (\boldsymbol{S} + \boldsymbol{A})^{-1} : (-\boldsymbol{\epsilon^0}), \qquad [30]$$

where $\boldsymbol{A}$ has been defined in eq. (10).

### 13. Expression of Eshelby's tensor $S$ for cylindrical inclusion

For cylindrical inclusion with elliptic face of semi axis length $a$ (along x-direction) and semi axis length $c$ (along z-direction), the components of Eshelby's tensor can be written as (13)

$$\boldsymbol{S}_{1111} = \frac{1}{2(1-\nu_b)}\left[\frac{c^2+2ac}{(a+c)^2} + (1-2\nu_b)\frac{c}{a+c}\right], \qquad [31]$$

$$\boldsymbol{S}_{2222} = \frac{1}{2(1-\nu_b)}\left[\frac{a^2+2ac}{(a+c)^2} + (1-2\nu_b)\frac{a}{a+c}\right], \qquad [32]$$

$$\boldsymbol{S}_{3333} = 0, \qquad [33]$$

$$\boldsymbol{S}_{1122} = \frac{1}{2(1-\nu_b)}\left[\frac{c^2}{(a+c)^2} - (1-2\nu_b)\frac{c}{a+c}\right], \qquad [34]$$

$$\boldsymbol{S}_{2233} = \frac{1}{2(1-\nu_b)}\frac{2\nu_b a}{a+c}, \qquad [35]$$

$$\boldsymbol{S}_{2211} = \frac{1}{2(1-\nu_b)}\left[\frac{a^2}{(a+c)^2} - (1-2\nu_b)\frac{a}{a+c}\right], \qquad [36]$$

$$\boldsymbol{S}_{3311} = 0, \boldsymbol{S}_{3322} = 0, \qquad [37]$$

$$\boldsymbol{S}_{1212} = \frac{1}{2(1-\nu_b)}\left[\frac{a^2+c^2}{(a+c)^2} + \frac{(1-2\nu_b)}{2}\right], \qquad [38]$$

$$\boldsymbol{S}_{1133} = \frac{1}{2(1-\nu_b)}\frac{2\nu_b c}{a+c}, \qquad [39]$$

$$\boldsymbol{S}_{2323} = \frac{a}{2(a+c)}, \qquad [40]$$

$$\boldsymbol{S}_{3131} = \frac{c}{2(a+c)}. \qquad [41]$$

Eshelby's tensor for spherical faced cylindrical inclusion can be found by setting $a = c$.

## 14. Expression of Eshelby's tensor $S$ for the flat ellipsoid-shaped inclusion

In case of flat shaped inclusion with elliptic face, assuming $a > b >> c$, the $I$-terms of eqs. 13-16 can be calculated as (13)

$$I_1 = 4\pi \left[\chi(k) - \zeta(k)\right] \frac{bc}{(a^2 - b^2)}, \tag{42}$$

$$I_2 = 4\pi \left[\frac{c}{b}\zeta(k) - \left(\chi(k) - \zeta(k)\right)\frac{bc}{(a^2 - b^2)}\right], \tag{43}$$

$$I_3 = 4\pi \left[1 - \frac{c}{b}\zeta(k)\right], \tag{44}$$

$$I_{12} = 4\pi \left[\frac{c}{b}\zeta(k) - 2\left(\chi(k) - \zeta(k)\right)\frac{bc}{(a^2 - b^2)}\right]/(a^2 - b^2), \tag{45}$$

$$I_{23} = 4\pi \left[1 - 2\frac{c}{b}\zeta(k) + \left(\chi(k) - \zeta(k)\right)\frac{bc}{(a^2 - b^2)}\right]/b^2, \tag{46}$$

$$I_{31} = 4\pi \left[1 - \frac{c}{b}\zeta(k) - \left(\chi(k) - \zeta(k)\right)\frac{bc}{(a^2 - b^2)}\right]/a^2, \tag{47}$$

$$I_{33} = \frac{4\pi}{3c^2}, \tag{48}$$

where the elliptic integrals $\chi$ and $\zeta$ are defined as

$$\chi(k) = \int_0^{\frac{\pi}{2}} \frac{dw}{\sqrt{1 - k^2 sin^2 w}} \tag{49}$$

$$\zeta(k) = \int_0^{\frac{\pi}{2}} \sqrt{1 - k^2 sin^2 w}\, dw. \tag{50}$$

Using the expressions of the I-terms determined above in eqs. 13-16, the Eshelby's tensor for the flat ellipsoid-shaped inclusion can be determined.

## 15. Expression of Eshelby's tensor $S$ for the penny-shaped inclusion

For penny-shaped inclusion with radius $a = c >> b$, the components of Eshelby's tensor can be written as (17, p. 81)

$$\boldsymbol{S}_{1111} = \boldsymbol{S}_{2222} = \frac{13 - 8\nu_b}{32(1 - \nu_b)} \pi \frac{b}{a}, \tag{51}$$

$$\boldsymbol{S}_{3333} = 1 - \frac{1 - 2\nu_b}{1 - \nu_b}\frac{\pi}{4}\frac{b}{a}, \tag{52}$$

$$\boldsymbol{S}_{1122} = \boldsymbol{S}_{2211} = \frac{8\nu_b - 1}{32(1 - \nu_b)} \pi \frac{b}{a}, \tag{53}$$

$$\boldsymbol{S}_{1133} = \boldsymbol{S}_{2233} = \frac{2\nu_b - 1}{8(1 - \nu_b)} \pi \frac{b}{a}, \tag{54}$$

$$\boldsymbol{S}_{3311} = \boldsymbol{S}_{3322} = \frac{\nu_b}{(1 - \nu_b)}\left(1 - \frac{4\nu_b + 1}{8\nu_b}\frac{b}{a}\right), \tag{55}$$

$$\boldsymbol{S}_{1212} = \frac{7 - 8\nu_b}{32(1 - \nu_b)} \pi \frac{b}{a}, \tag{56}$$

$$\boldsymbol{S}_{1313} = \boldsymbol{S}_{2323} = \frac{1}{2}\left(1 + \frac{\nu_b - 2}{1 - \nu_b}\frac{\pi}{4}\frac{b}{a}\right). \tag{57}$$

## 16. Expressions of $\epsilon_1^*$ and $\epsilon_2^*$ for spherical inclusion

For spherical inclusion (tumor), the Eshelby tensor's components can be written as (15, 16)

$$S_{1111} = S_{2222} = S_{3333} = m_1 = \frac{7 - 5\nu_b}{15(1 - \nu_b)}, \qquad [58]$$

$$S_{1122} = S_{2233} = S_{3311} = S_{2211} = S_{3322} = S_{1133} = m_2 = \frac{5\nu_b - 1}{15(1 - \nu_b)}, \qquad [59]$$

$$S_{1212} = S_{2323} = S_{3131} = m_3 = \frac{4 - 5\nu_b}{15(1 - \nu_b)}. \qquad [60]$$

The eigen strain can be written as (4, 14)

$$\boldsymbol{\epsilon}_1^* = \boldsymbol{S}^{-1}[\boldsymbol{\epsilon} - \boldsymbol{\epsilon^0}]. \qquad [61]$$

Using $\epsilon_{11} = \epsilon_{33}, \epsilon_{11}^0 = \epsilon_{33}^0$ for spherical inclusion, the first expression of $\epsilon_1^*$ can be written as

$$\boldsymbol{\epsilon}_1^* = \begin{bmatrix} \frac{m_1(\epsilon_{11} - \epsilon_{11}^0)}{m_1^2 + m_1 m_2 - 2m_2^2} - \frac{m_2(\epsilon_{22} - \epsilon_{22}^0)}{m_1^2 + m_1 m_2 - 2m_2^2} \\ \frac{(m_1 + m_2)(\epsilon_{22} - \epsilon_{22}^0)}{m_1^2 + m_1 m_2 - 2m_2^2} - \frac{2m_2(\epsilon_{11} - \epsilon_{11}^0)}{m_1^2 + m_1 m_2 - 2m_2^2} \\ \frac{m_1(\epsilon_{11} - \epsilon_{11}^0)}{m_1^2 + m_1 m_2 - 2m_2^2} - \frac{m_2(\epsilon_{22} - \epsilon_{22}^0)}{m_1^2 + m_1 m_2 - 2m_2^2} \\ 0 \\ 0 \\ 0 \end{bmatrix}. \qquad [62]$$

Another expression of the eigen strain can be written as (15, 16)

$$\boldsymbol{\epsilon}_2^* = (\boldsymbol{S} + \boldsymbol{A})^{-1} : (-\boldsymbol{\epsilon^0}). \qquad [63]$$

Let us now expand this equation in terms of the YM and PR of the inclusion and background.

From eq. 63, $\boldsymbol{H} = (\boldsymbol{S} + \boldsymbol{A})^{-1}$ can be written as

$$\boldsymbol{H} = \begin{bmatrix} D_1 & D_2 & D_2 & 0 & 0 & 0 \\ D_2 & D_1 & D_2 & 0 & 0 & 0 \\ D_2 & D_2 & D_1 & 0 & 0 & 0 \\ 0 & 0 & 0 & 1/(a_3 + m_3) & 0 & 0 \\ 0 & 0 & 0 & 0 & 1/(a_3 + m_3) & 0 \\ 0 & 0 & 0 & 0 & 0 & 1/(a_3 + m_3) \end{bmatrix}, \qquad [64]$$

where

$$D_1 = (a_1 + a_2 + m_1 + m_2)/(a_1^2 + a_1 a_2 + 2a_1 m_1 + a_1 m_2 - 2a_2^2 + a_2 m_1 - 4a_2 m_2 + m_1^2 + m_1 m_2 - 2m_2^2), \qquad [65]$$

$$D_2 = -(a_2 + m_2)/(a_1^2 + a_1 a_2 + 2a_1 m_1 + a_1 m_2 - 2a_2^2 + a_2 m_1 - 4a_2 m_2 + m_1^2 + m_1 m_2 - 2m_2^2). \qquad [66]$$

Here

$$a_1 = \Psi(\lambda_b + 2\mu_b) + 2\lambda_b \Omega, \qquad [67]$$

$$a_2 = \Omega(\Lambda_b + 2\mu_b) + \lambda_b \Omega + \lambda_b \Psi, \qquad [68]$$

$$a_3 = \mu_b \Phi \qquad [69]$$

and

$$\Psi = -(\lambda_b - \lambda_i + \mu_b - \mu_i)/(2\mu_b^2 + 2\mu_i^2 + 3\lambda_b \mu_b - 3\lambda_b \mu_i - 3\lambda_i \mu_b + 3\lambda_i \mu_i - 4\mu_b \mu_i), \qquad [70]$$

$$\Omega = (\lambda_b - \lambda_i)/(2(2\mu_b^2 + 2\mu_i^2 + 3\lambda_b \mu_b - 3\lambda_b \mu_i - 3\lambda_i \mu_b + 3\lambda_i \mu_i - 4\mu_b \mu_i)), \qquad [71]$$

$$\Phi = -\frac{1}{\mu_b - \mu_i}. \qquad [72]$$

The expression of $\epsilon_2^*$ can now be obtained as

$$\boldsymbol{\epsilon}_2^* = \begin{bmatrix} -\frac{\epsilon_{11}^0(a_1 + m_1)}{L} + \frac{\epsilon_{22}^0(a_2 + m_2)}{L} \\ \frac{2\epsilon_{11}^0(a_2 + m_2)}{L} - \frac{\epsilon_{22}^0(a_1 + a_2 + m_1 + m_2)}{L} \\ -\frac{\epsilon_{11}^0(a_1 + m_1)}{L} + \frac{\epsilon_{22}^0(a_2 + m_2)}{L} \\ 0 \\ 0 \\ 0 \end{bmatrix}, \qquad [73]$$

where

$$L = a_1^2 + a_1 a_2 + 2a_1 m_1 + a_1 m_2 - 2a_2^2 + a_2 m_1 - 4a_2 m_2 + m_1^2 + m_1 m_2 - 2m_2^2. \qquad [74]$$

## 17. Readings from the force sensor in experiments in vivo

Readings from the force sensor in an in vivo elastography experiment is shown in Fig. S18.

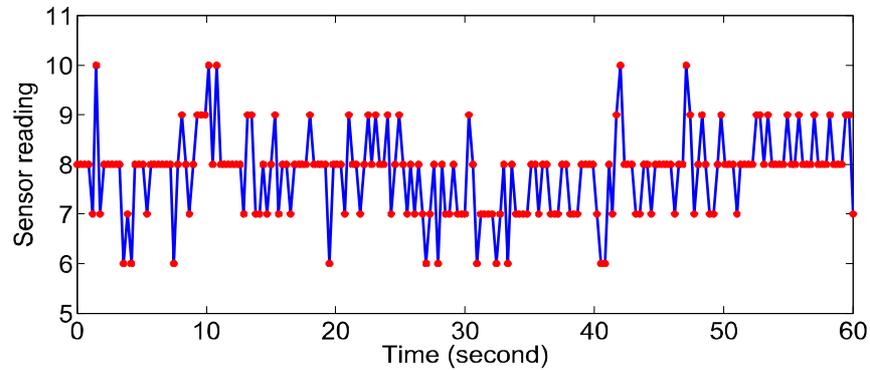

**Fig. S18.** Reading from the force sensor in an in vivo elastography experiment.

## 18. Approximation of different shapes with ellipses

In the proposed approach of YM and PR reconstruction, the complex shapes such as tetragon, pentagon and hexagon are approximated with ellipses. These shapes along with the approximated ellipses are shown in Fig. S19.

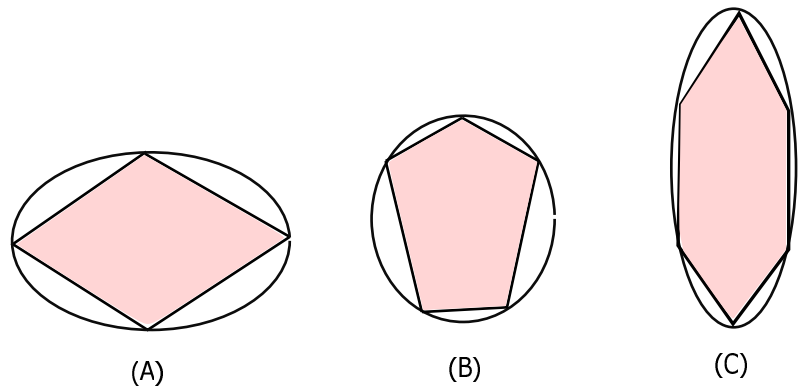

**Fig. S19.** Approximation of different shapes with ellipses (A) tetragon (B) pentagon (C) hexagon. It has been assumed that in the imaging plane the tumors are of these shapes and if the plane is revolved around the center line, the shape remains the same in all other planes (axisymmetry).

## 19. Three dimensional samples

**A. Cubic samples with elliptical inclusion.** We choose three cubic samples (V1-V3) with elliptical inclusion. The three dimensional cubic samples along with the solution space used in the FEA are shown in Fig. S20. The size of the solution space is 2 cm along all three directions. The mesh element has been used in all three dimensional samples is C3D8P. The number of mesh element in the solution space of V1-V3 is 133067. In all the samples, the YM and PR of the normal tissue has been taken as 32.78 kPa and 0.4. The YM and PR of the inclusion has been taken as 97.02 kPa and 0.4. The size of all three dimensional cubic samples is assumed $4 \times 4 \times 4$ cm$^3$. The lengths of the semi-axes of the inclusions in each sample has been shown in Table S3. The RMSEs in reconstructed YM and PR by the proposed technique with and without axisymmetric assumption in samples V1-V3 are shown in Table S4.

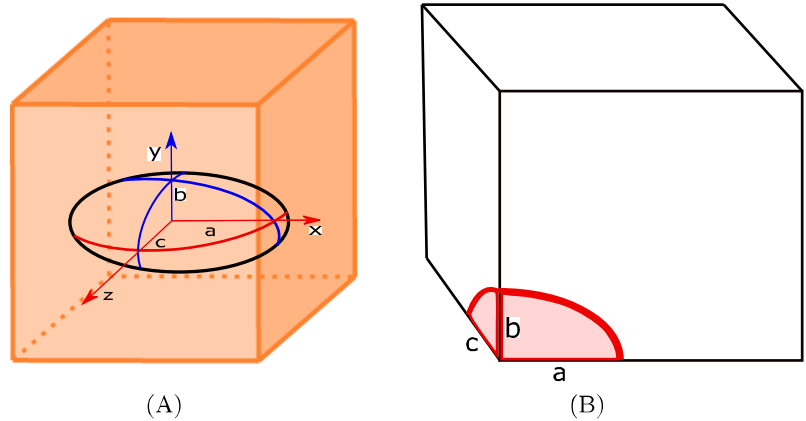

**Fig. S20.** (A) A cubic sample of a poroelastic material with an elliptical poroelastic inclusion of semi-axes $a, b$ and $c$ along $x, y$ and $z$ direction. The compression is along the negative $y$-direction. (B) Solution space.

**Table S3. Size of the inclusions in samples V1-V3 used in three-dimensional FE simulations**

| Sample name | a (cm) | b (cm) | c (cm) |
|---|---|---|---|
| V1 | 0.5 | 0.25 | 0.35 |
| V2 | 0.3 | 0.9 | 0.3 |
| V3 | 0.25 | 0.35 | 0.45 |

**Table S4. RMSE (%) in three-dimensional reconstruction and reconstruction by axisymmetric assumption**

| Sample | 3D reconstruction | | axisymmetric assumption | |
|---|---|---|---|---|
| | YM | PR | YM | PR |
| V1 | 2.80 | 4.07 | 3.10 | 4.76 |
| V2 | 3.18 | 4.01 | 5.04 | 4.13 |
| V3 | 3.51 | 4.12 | 5.15 | 4.23 |

**B. Samples in plane stress and plane strain.** Plane strain and plane stress are conditions that may arise in the three-dimensional analysis of the elastic behavior of materials but are of limited relevance for the analysis of tissues. More specifically, plane stress conditions may occur in a sample that is very thin along the z-direction, which is compressed from the x or y direction. In such situation, the stress along the z-direction may be assumed to be zero. On the other hand, plane strain conditions may occur in a cylindrical sample that is infinitely long in the z-direction, and the applied stress is along the periphery of the sample. In such situation, the strain along the z-direction may be assumed to be zero. In elastography applications, plane stress and plane strain conditions are rarely satisfied as most samples have lengths of the same order of magnitude along the three directions. In samples with long cylindrical inclusions such as those considered in Refs. (10, 18, 19), the assumption of plane strain inside the inclusion may be justified. However, these cases represent idealized situations that rarely present in experimental applications and have limited relevance in the analysis of the behavior of tumors and biological tissues, in general.

We simulate three-dimensional samples with inclusion in plane stress and plane strain and reconstruct the YM and PR of them by the proposed method. We describe the procedure of simulation and associated results in the following subsections.

**B.1. Thin elliptical inclusion (plane stress).** We simulate two cubic samples (K1 and K2) with ellipsoidal inclusions of very small thickness along z-direction ($a = 0.3, b = 0.3$ and $c = 0.04$ cm). The number of mesh element in solution space of K1-K2 is 216264. In sample K1, the YM and PR of the inclusion has been taken as 97.02 kPa and 0.4, whereas in sample K2, the YM and PR of the inclusion has been considered 163.90 kPa and 0.4. The YM and PR of the normal tissue is considered as 32.78 kPa and 0.4 in both samples. We show the elevational stress for this sample in Fig. S21 for an applied load of 1 kPa. Very small value of elevational stress inside the inclusion proves that the inclusion is indeed in plane stress condition.

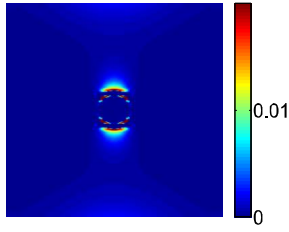

**Fig. S21.** Elevational stress (kPa) in 3D sample with thin inclusion, which imitates a plane stress condition

**Eshleby's tensor**   For an inclusion with radius $a = b \gg c$, which imitates the plane stress condition of the inclusion, the components of Eshleby's tensor can be written as (17, p. 81)

$$S_{1111} = S_{2222} = \frac{13 - 8\nu_b}{32(1 - \nu_b)} \pi \frac{c}{a}, \tag{75}$$

$$S_{3333} = 1 - \frac{1 - 2\nu_b}{1 - \nu_b} \frac{\pi}{4} \frac{c}{a}, \tag{76}$$

$$S_{1122} = S_{2211} = \frac{8\nu_b - 1}{32(1 - \nu_b)} \pi \frac{c}{a}, \tag{77}$$

$$S_{1133} = S_{2233} = \frac{2\nu_b - 1}{8(1 - \nu_b)} \pi \frac{c}{a}, \tag{78}$$

$$S_{3311} = S_{3322} = \frac{\nu_b}{(1 - \nu_b)} \left(1 - \frac{4\nu_b + 1}{8\nu_b} \frac{c}{a}\right), \tag{79}$$

$$S_{1212} = \frac{7 - 8\nu_b}{32(1 - \nu_b)} \pi \frac{c}{a}, \tag{80}$$

$$S_{1313} = S_{2323} = \frac{1}{2}\left(1 + \frac{\nu_b - 2}{1 - \nu_b} \frac{\pi}{4} \frac{c}{a}\right). \tag{81}$$

**RMSE in estimated YM and PR**   RMSE values in estimated YM and PR of K1 and K2 by the proposed technique are tabulated in Table S5.

**Table S5. RMSE (%) in estimation of YM and PR of K1-K2**

| Sample name | YM | PR |
|---|---|---|
| K1 | 5.17 | 0.18 |
| K2 | 6.15 | 1.19 |

***B.2. Infinitely long cylindrical inclusion (plane strain).*** We simulate two cubic samples (J1 and J2) with infinitely long cylindrical inclusion along $z$-direction ($a = 0.3, b = 0.3$ cm) as shown in Fig. S22, which imitates the plane strain condition. The number of mesh element in solution space of J1-J2 is 142532. We apply the boundary condition of zero elevational displacement along all the surfaces of the inclusion also to make sure that there is no elevational strain in the inclusion. In sample J1, the YM and PR of the inclusion has been taken as 97.02 kPa and 0.4, whereas in sample J2, the YM and PR of the inclusion has been considered 163.90 kPa and 0.4. The YM and PR of the normal tissue is considered as 32.78 kPa and 0.4 in both samples. We show the elevational strain for this sample in Fig. S23. Very small value of elevational strain inside the inclusion proves that the inclusion is indeed in plane strain condition.

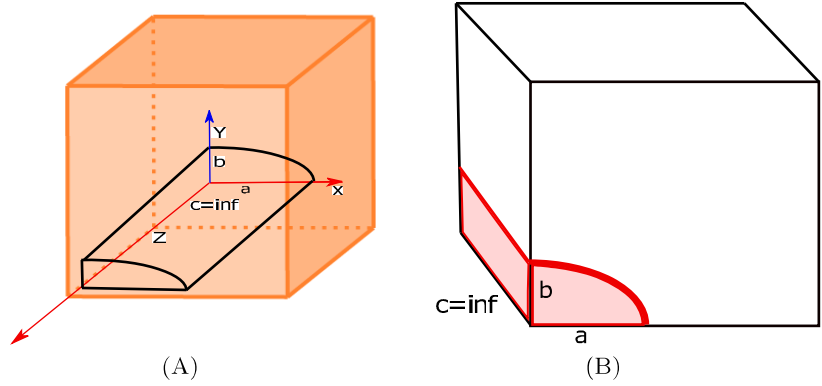

**Fig. S22.** (A) A schematic of a cubic sample of a poroelastic sample with a cylindrical poroelastic inclusion of elliptical face. (B) Solution space

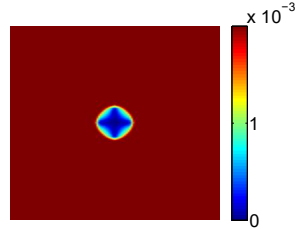

**Fig. S23.** Elevational strain in a 3D sample with long cylindrical inclusion, which imitates a plane strain condition

**Eshleby's tensor**  For an inclusion with elliptic face of semi-axis length $a$ along x-axis and semi-axis length $b$ along y-axis in plane strain state, the components of Eshleby's tensor can be written as (13)

$$S_{1111} = \frac{1}{2(1-\nu_b)} \left[ \frac{b^2 + 2ab}{(a+b)^2} + (1-2\nu_b)\frac{b}{a+b} \right], \quad [82]$$

$$S_{2222} = \frac{1}{2(1-\nu_b)} \left[ \frac{a^2 + 2ab}{(a+b)^2} + (1-2\nu_b)\frac{a}{a+b} \right], \quad [83]$$

$$S_{3333} = 0, \quad [84]$$

$$S_{1122} = \frac{1}{2(1-\nu_b)} \left[ \frac{b^2}{(a+b)^2} - (1-2\nu_b)\frac{b}{a+b} \right], \quad [85]$$

$$S_{2233} = \frac{1}{2(1-\nu_b)} \frac{2\nu_b a}{a+b}, \quad [86]$$

$$S_{2211} = \frac{1}{2(1-\nu_b)} \left[ \frac{a^2}{(a+b)^2} - (1-2\nu_b)\frac{a}{a+b} \right], \quad [87]$$

$$S_{3311} = 0, S_{3322} = 0, \quad [88]$$

$$S_{1212} = \frac{1}{2(1-\nu_b)} \left[ \frac{a^2 + b^2}{(a+b)^2} + \frac{(1-2\nu_b)}{2} \right], \quad [89]$$

$$S_{1133} = \frac{1}{2(1-\nu_b)} \frac{2\nu_b b}{a+b}, \quad [90]$$

$$S_{2323} = \frac{a}{2(a+b)}, \quad [91]$$

$$S_{3131} = \frac{b}{2(a+b)}. \quad [92]$$

**RMSE in the estimated YM and PR** tabulated in Table S6.  RMSE values in the estimated YM and PR of J1 and J2 by the proposed technique are

**Table S6. RMSE (%) in the estimation of YM and PR of J1-J2**

| Sample name | YM   | PR   |
|:-----------:|:----:|:----:|
| J1          | 3.67 | 1.32 |
| J2          | 5.82 | 2.18 |

**C. Samples with multiple inclusions.** We simulate four cubic samples (U1-U4) with three inclusions inside it as shown in Figs. S24 and S25. The central, right (top) and left (bottom) inclusions are numbered as 1, 2 and 3. All inclusions are assumed spherical. The radius of inclusion 1 is assumed $a_1 = 0.3$ cm and of inclusion 2 and 3 is assumed $a_2 = 0.2$ cm. The distance between inclusion 1 to other two inclusions is $d = 0.65$ cm. The material properties of the inclusions of U1-U4 are tabulated in Table S7. The YM of the background is taken as 32.78 kPa in all the samples. The PR of the background of U1 and U3 is taken as 0.4 and of U2 and U4 is taken as 0.3. The RMSEs in the reconstructed YM and PR by the proposed method in the inclusions in these samples are tabulated in Table S8.

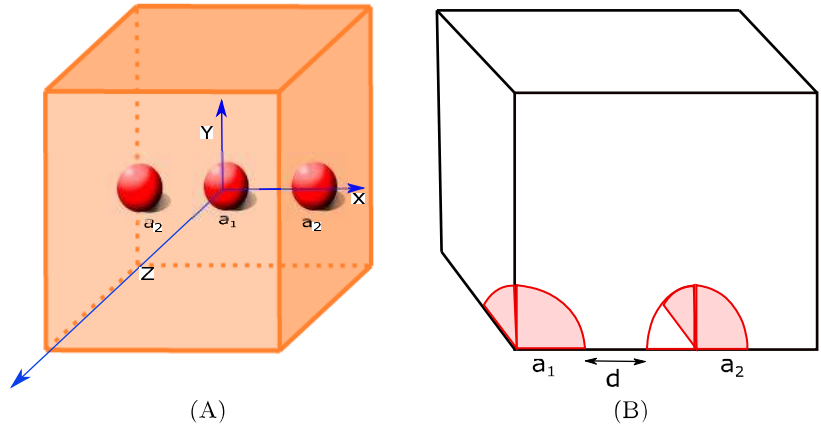

**Fig. S24.** (A) A schematic of a cubic sample (U1-U2) of a poroelastic material with multiple spherical poroelastic inclusions side by side. The axial direction is along the $y$-axis. (B) Solution space

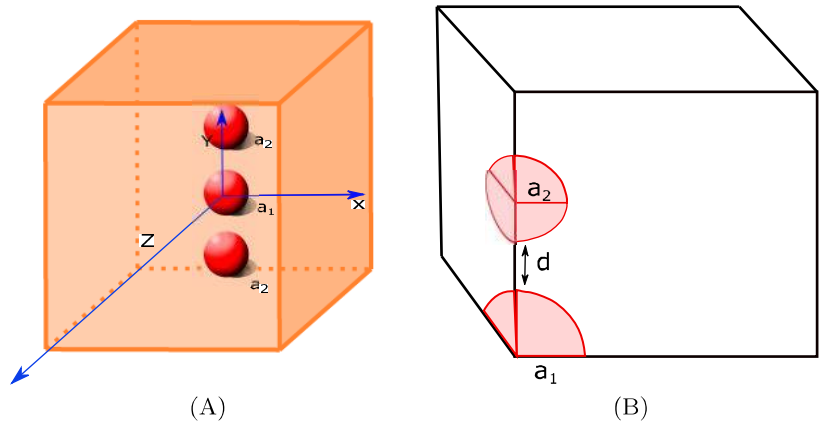

**Fig. S25.** (A) A schematic of a cubic sample (U3-U4) of a poroelastic material with multiple spherical poroelastic inclusions on top of each other. The axial direction is along the $y$-axis. (B) Solution space

Table S7. Material properties of U1-U4

| Sample | inclusion 1 | | inclusion 2 | | inclusion 3 | |
|---|---|---|---|---|---|---|
| | YM (kPa) | PR | YM (kPa) | PR | YM (kPa) | PR |
| U1 | 97.02 | 0.4 | 65.56 | 0.4 | 65.56 | 0.4 |
| U2 | 163.90 | 0.3 | 81.95 | 0.3 | 81.95 | 0.3 |
| U3 | 97.02 | 0.4 | 65.56 | 0.4 | 65.56 | 0.4 |
| U4 | 163.90 | 0.3 | 81.95 | 0.3 | 81.95 | 0.3 |

Table S8. RMSE (%) in estimation of YM and PR of U1-U4

| Sample | inclusion 1 | | inclusion 2 | | inclusion 3 | |
|---|---|---|---|---|---|---|
| | YM | PR | YM | PR | YM | PR |
| U1 | 0.7 | 1.07 | 4.39 | 3.80 | 5.49 | 2.17 |
| U2 | 0.6 | 1.33 | 7.67 | 8.00 | 8.54 | 5.67 |
| U3 | 10.96 | 0.51 | 0.81 | 0.01 | 3.33 | 0.27 |
| U4 | 10.11 | 0.72 | 0.62 | 0.02 | 2.83 | 0.21 |

## 20. Results of additional controlled experiments

**Table S9. Mean and standard deviation of the reconstructed YM and PR distributions in controlled experiments**

| Exp no | Applied load (kPa) | Est YM of inclusion (kPa) | Est PR of inclusion | Est YM of background (kPa) | Est PR of background |
|---|---|---|---|---|---|
| CE2 | 1.27 | 50.60 ±5.45 | 0.45 ±0.01 | 21.02 ±8.67 | 0.42 ±0.03 |
| CE3 | 1.48 | 46.21 ±4.16 | 0.47 ±0.03 | 18.53 ±4.19 | 0.42 ±0.02 |
| CE4 | 1.16 | 47.73 ±3.93 | 0.44 ±0.01 | 17.77 ±2.71 | 0.41 ±0.05 |
| CE5 | 1.97 | 49.46 ±3.28 | 0.47 ±0.01 | 21.88 ±9.13 | 0.44 ±0.06 |

## 21. YM validation using a shear wave imaging system

The breast phantom used for the controlled experiments is a linear elastic phantom made of incompressible material (as indicated by the manufacturer). Therefore, the value of PR in both the inclusion and background is 0.5. To validate our reconstructed YM results obtained from the breast phantom using an independent non-destructive methodology, a shear wave elastography system was used (20). The YM image estimated using the shear wave elastography system is shown in Fig. S26. From the shear wave data, the mean YM was estimated to be 42.725 kPa in the inclusion and 18.15 kPa in the background. These values closely match ($< 9\%$ difference) the YM results obtained using our proposed reconstruction method. We could not directly validate the reconstructed PR image using an independent technique since there are no imaging methods retrievable in the literature that can assess the PR. However, our estimated PR values match well ($<15\%$ difference) those reported by the manufacturer of this phantom, a result that in itself can serve as a validation of the accuracy of the proposed method.

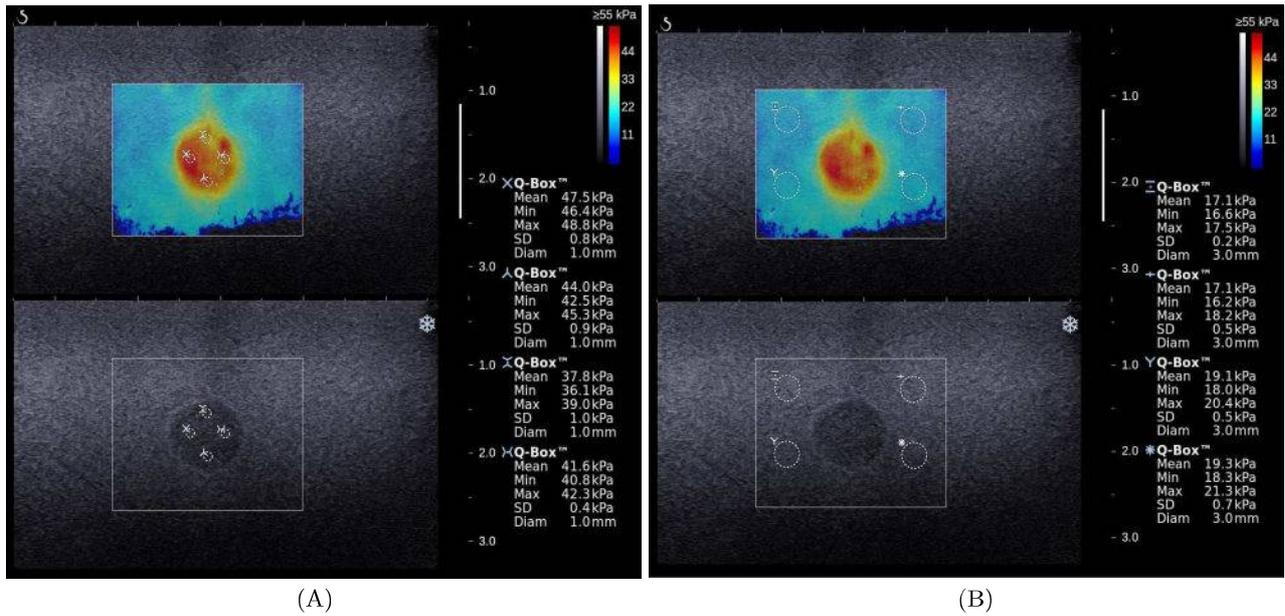

(A)          (B)

**Fig. S26.** Measurement of YM using a shear wave imaging system: (A) inside the inclusion and (B) inside the background.

## 22. Validation of local stress measurement

Table S10. Mean and standard deviation of the axial, lateral and shear stresses inside the inclusion

| Sample | FEA | | | Proposed method | | |
|---|---|---|---|---|---|---|
| | Axial stress (kPa) | Lateral stress (kPa) | Shear stress (kPa) | Axial stress (kPa) | Lateral stress (kPa) | Shear stress (kPa) |
| A | 1.435 ±0.011 | -0.249 ±0.016 | 0.0033 ±0.0003 | 1.437 ±0.001 | -0.242 ±0.0006 | 0 |
| B | 1.153 ±0.002 | -0.140 ±0.004 | 0.0001 ±0.001 | 1.153 ±0.018 | -0.139 ±0.0005 | 0 |
| C | 1.608 ±0.008 | -0.319 ±0.012 | -0.001 ±0.004 | 1.614 ±0.001 | -0.312 ±0.0006 | 0 |
| D | 1.435 ±0.001 | -0.174 ±0.005 | 0.0008 ±0.001 | 1.44 ±0.003 | -0.174±0.001 | 0 |

In the background region, in both FEA and the proposed method, the axial stress is equal to the applied stress and the lateral and shear stresses are zero.

## 23. Creep experiment and determination of steady state axial and lateral strains

We used a creep experiment to estimate the YM and PR in the tumors and normal tissues (21, 22). In such experiments, immediately after the compression is applied, the fluid pressure inside the compressed region of the tissue (both tumor and background) rises, and the tissue behaves as an incompressible material with an effective Poisson's ratio (EPR) of 0.5. As the fluid pressure at the outer boundary of the compressed region of the tissue is zero, for $t > 0$ $s$, a fluid pressure gradient is created. Another fluid pressure gradient is created because of the lower pressure on the other ends of the capillary vessels inside normal tissue and tumor. The pressure gradients cause fluid exudation and facilitate fluid flow toward the boundary of the compressed region of the tissue and through the capillary vessels. The lateral stress in the solid matrix of the tissue acts along the opposite direction of the fluid flow. When the fluid flow ceases and the solid matrix of the tumor and background are fully relaxed, the EPR of the tumor and background become the Poisson's ratio. At this point, the tissue (both tumor and background) behave as linearly elastic material.

The axial and lateral strains that generate in the tissue under compression are time-dependent, due to fluid translocation effects. Steady state conditions occur when the tissue reaches equilibrium and fluid flow ceases. At steady state, the strains no longer vary with time. Therefore, by analyzing the strain profiles, it is possible to detect when steady state conditions occur. Moreover, this can be done automatically. To illustrate this point, in Fig. S27, we plot the axial and lateral strains inside one untreated tumor. We see that, after 35 s, the axial and lateral strains do not change with time. Thus, in this example, the average value of the strain from 35 to 60 s can be used as the steady state value to reconstruct the YM and PR.

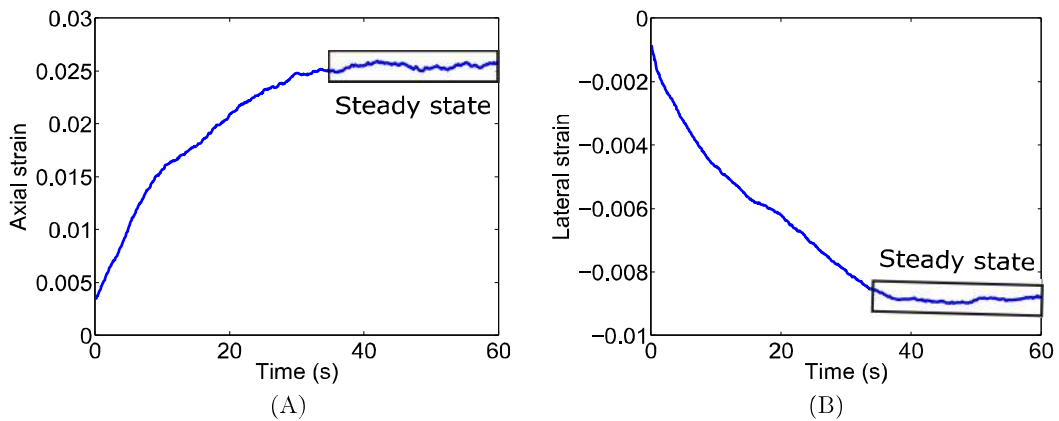

**Fig. S27.** Creep curves of axial and lateral strains.

## 24. Experimental setup for elastography experiments on breast phantom

The experimental setup for the controlled elastography experiments on the breast phantom is shown in Fig. S28. Different components of the experimental setup are shown in this figure.

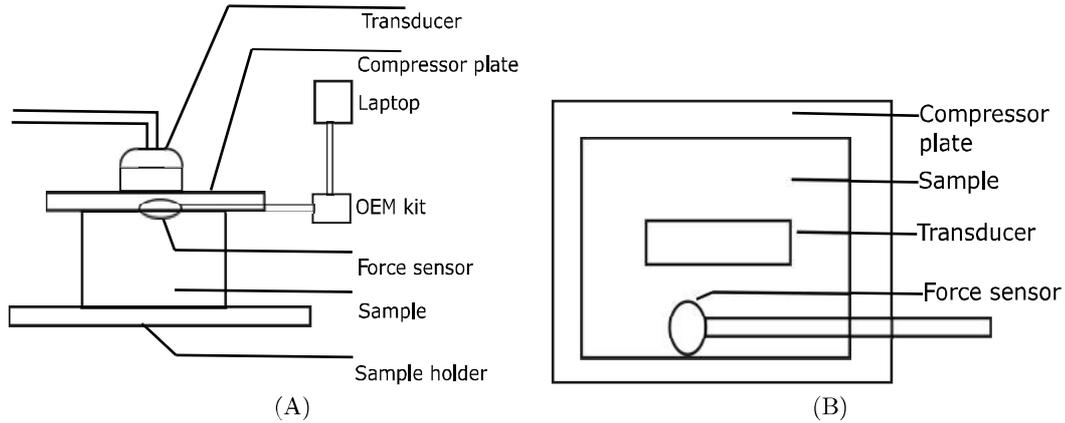

**Fig. S28.** (A) Experimental setup (B) Top view of the setup showing the placement of the force sensor between the compressor plate and sample

## 25. Effect of spatial and temporal non-uniformity of applied stress on reconstructed YM and PR by proposed method

In our experiments, we used a small force sensor so that it would not interfere with the ultrasonic data acquisition. Since the sensor is small and placed in a fixed location, it does not ensure that the applied stress is uniformly distributed over the entire sample. However, the tissue was compressed using a compressor plate of size 10.1 cm × 9.3 cm, i.e., large with respect to the size of the inclusions. In section 8 of the supplementary material, we prove that the use of a compressor plate larger than the tissue region of interest guarantees a stress distribution in the imaging plane that does not spatially change in a significant manner. This has also been proven theoretically in Ophir et al. (12).

While, theoretically, the compressor plate should assure a uniform applied stress, the spatial distribution of the stress may not be completely uniform because of operator's motion, breathing motion from the animals etc. Thus, we have also analyzed the effects of non-uniform loading on our YM and PR reconstruction using simulations and reported the results in Table 5. We found out that, even if the loading is increased or reduced by 20% along the compression axis, our method can still reconstruct the YM with more than 90% and the PR with more than 96% accuracy.

Moreover, since the experimental data is collected in free-hand mode, the applied stress may not be uniform within the 1 minute acquisition interval. This can also be observed in the added temporal profiles of the applied force obtained using the force sensor (see Fig. S18). We report the mean and standard deviation of the applied stress in four in vivo experiments in Table S11. We see that the standard deviations of the applied compression are $< 10\%$ in all cases. Such small deviations in the applied stress would not affect the accuracy of the reconstructed YM and PR by our method significantly as proven by the results shown in Table 5.

**Table S11. Mean and standard deviation of the applied force**

| In vivo experiments | Mean (Pa) | Standard deviation (Pa) |
|---|---|---|
| 1 | 1891 | 120 |
| 2 | 2242 | 197 |
| 3 | 2413 | 159 |
| 4 | 2357 | 214 |

## 26. Effect of out of plane and breathing motion on reconstructed YM and PR by proposed method

In general, out of plane motion can affect the accuracy of strain estimation and as such the quality of the reconstructed YM and PR. When a tissue is compressed, the motion of the scatterers inside the tissue is three-dimensional, and, in general, the elevational displacement of these scatterers cannot be properly estimated using linear array probes. If the compression is very large, out of plane motion can significantly affect the displacement and strain estimations. The 'Strain Filter' theory proposed by Varghese and Ophir predicts these high compression strain levels above which decorrelation significantly affects the elastographic signal-to-noise ratio (23). In our experiments, we use the creep-compression approach typically used in poroelastography experiments in vivo so that the strain between successive frames would always be below 0.1-0.2% (24). At these compression levels, decorrelation noise due to out of plane motion is typically negligible (23). Moreover, our strain estimation method proposed in Islam et al. (8) assumes similarity of echo amplitudes and displacement continuity while estimating the strains, which further reduces the noise due to decorrelation (25).

Motion due to breathing is a periodic motion (26), which can be source of decorrelation noise in the strain elastograms. Breathing artifacts can be minimized by using high frame rate ultrasound systems and by computing the displacements and strains from successive RF frames or RF frames separated by sufficiently short time intervals.

## 27. Computational requirements of the proposed method

As the proposed method to estimate the YM and PR requires optimization of a cost function in every pixel inside the tumor, the computation time for the proposed technique is higher in comparison to some of the previously methods. In the present configuration, the simultaneous reconstruction of YM and PR requires 1.7 s on average in an Intel Xeon 3.5 GHz PC with 32 GB RAM for each pixel inside the tumor and less than 1 s for all the pixels outside the tumor (with FE simulation axial and lateral strain data for the computational setup described in the Methods section), whereas the methods in Refs. (14) (9) require less than 1 s for the entire image.

1. Islam MT, Chaudhry A, Unnikrishnan G, Reddy J, Righetti R (2018) An analytical poroelastic model for ultrasound elastography imaging of tumors. *Physics in Medicine & Biology* 63(2):025031.
2. Leiderman R, Barbone PE, Oberai AA, Bamber JC (2006) Coupling between elastic strain and interstitial fluid flow: ramifications for poroelastic imaging. *Physics in medicine and biology* 51(24):6291.
3. (2017) Detecting a cell using image segmentation. Accessed: 2017-08-20.
4. Eshelby JD (1957) The determination of the elastic field of an ellipsoidal inclusion, and related problems in *Proceedings of the Royal Society of London A: Mathematical, Physical and Engineering Sciences*. (The Royal Society), Vol. 241, pp. 376–396.
5. Desai RR, Krouskop TA, Righetti R (2010) Elastography using harmonic ultrasonic imaging: a feasibility study. *Ultrasonic imaging* 32(2):103–117.
6. Chaudhry A (2010) Master's thesis (Texas A&M University).
7. Righetti R, Ophir J, Srinivasan S, Krouskop TA (2004) The feasibility of using elastography for imaging the poisson's ratio in porous media. *Ultrasound in medicine & biology* 30(2):215–228.
8. Islam MT, Chaudhry A, Tang S, Tasciotti E, Righetti R (2018) A new method for estimating the effective poisson's ratio in ultrasound poroelastography. *IEEE Transactions on Medical Imaging*.
9. Bilgen M, Insana MF (1998) Elastostatics of a spherical inclusion in homogeneous biological media. *Physics in Medicine and Biology* 43(1):1.
10. Kallel F, Bertrand M, Ophir J (1996) Fundamental limitations on the contrast-transfer efficiency in elastography: an analytic study. *Ultrasound in medicine & biology* 22(4):463–470.
11. Ophir J, et al. (1996) Elastography: ultrasonic imaging of tissue strain and elastic modulus in vivo. *European journal of ultrasound* 3(1):49–70.
12. Ophir J, Cespedes I, Ponnekanti H, Yazdi Y, Li X (1991) Elastography: a quantitative method for imaging the elasticity of biological tissues. *Ultrasonic imaging* 13(2):111–134.
13. Weinberger C, Cai W (2004) Lecture note 2. eshelby's inclusion i.
14. Shin B, Gopaul D, Fienberg S, Kwon HJ (2016) Application of eshelby's solution to elastography for diagnosis of breast cancer. *Ultrasonic imaging* 38(2):115–136.
15. Ju J, Sun L (2001) Effective elastoplastic behavior of metal matrix composites containing randomly located aligned spheroidal inhomogeneities. part i: micromechanics-based formulation. *International Journal of Solids and Structures* 38(2):183–201.
16. Ju J, Sun L (1999) A novel formulation for the exterior-point eshelby's tensor of an ellipsoidal inclusion. *Journal of Applied Mechanics* 66(2):570–574.
17. Mura T (1987) Micromechanics of defects in solids. mechanics of elastic and inelastic solids, vol. 3.
18. Richards MS, Barbone PE, Oberai AA (2009) Quantitative three-dimensional elasticity imaging from quasi-static deformation: a phantom study. *Physics in Medicine & Biology* 54(3):757.
19. Doyley M, Meaney P, Bamber J (2000) Evaluation of an iterative reconstruction method for quantitative elastography. *Physics in medicine and biology* 45(6):1521.
20. Bercoff J, Tanter M, Fink M (2004) Supersonic shear imaging: a new technique for soft tissue elasticity mapping. *IEEE transactions on ultrasonics, ferroelectrics, and frequency control* 51(4):396–409.
21. Islam MT, Reddy J, Righetti A, Raffaella (2018) An analytical poroelastic model of a non-homogeneous medium under creep compression for ultrasound poroelastography applications - part i. *In press, Journal of Biomechanical Engineering*.
22. Islam MT, Reddy J, Righetti A, Raffaella (2018) An analytical poroelastic model of a non-homogeneous medium under creep compression for ultrasound poroelastography applications - part ii. *In press, Journal of Biomechanical Engineering*.
23. Varghese T, Ophir J (1997) A theoretical framework for performance characterization of elastography: The strain filter. *IEEE Transactions on Ultrasonics, Ferroelectrics, and Frequency Control* 44(1):164–172.
24. Righetti R, Ophir J, Garra BS, Chandrasekhar RM, Krouskop TA (2005) A new method for generating poroelastograms in noisy environments. *Ultrasonic imaging* 27(4):201–220.
25. Rivaz H, et al. (2008) Ultrasound elastography: a dynamic programming approach. *IEEE transactions on medical imaging* 27(10):1373–1377.
26. Foroughi P, et al. (2013) A freehand ultrasound elastography system with tracking for in vivo applications. *Ultrasound in Medicine and Biology* 39(2):211–225.